\documentclass[10pt, twocolumn, nofootinbib,superscriptaddress]{revtex4-1}
\usepackage{amsmath,amssymb,amsfonts}
\usepackage{algorithmic}
\usepackage{graphicx}
\usepackage{textcomp}
\usepackage{xcolor}
\usepackage{ragged2e}
\usepackage{booktabs, makecell, tabularx}

\usepackage{gensymb}
\makeatletter
    \renewcommand\@make@capt@title[2]{%
     \@ifx@empty\float@link{\@firstofone}{\expandafter\href\expandafter{\float@link}}%
      {\textbf{#1}}\@caption@fignum@sep#2\quad}%
\makeatother
\makeatletter 
\renewcommand{\fnum@figure}{\textbf{Fig.~\thefigure}}
\makeatother

\usepackage{xcolor}
\newcommand{\beginsupplement}{%
        \setcounter{table}{0}
        \renewcommand{\thetable}{S\arabic{table}}%
        \setcounter{figure}{0}
        \renewcommand{\thefigure}{S\arabic{figure}}%
     }
\def\BibTeX{{\rm B\kern-.05em{\sc i\kern-.025em b}\kern-.08em
    T\kern-.1667em\lower.7ex\hbox{E}\kern-.125emX}}

\newcommand{\RomanNumeralCaps}[1]{\MakeUppercase{\romannumeral #1}}

\begin{document}

\title{Ultrahigh Dynamic Range and Low Noise Figure Programmable Integrated Microwave Photonic Filter}

\author{Okky Daulay}
\thanks{These authors contributed equally}
\affiliation{Nonlinear Nanophotonics Group, MESA+ Institute of Nanotechnology
	University of Twente, Enschede, Netherlands}
\author{Gaojian Liu}
\thanks{These authors contributed equally}
\affiliation{Nonlinear Nanophotonics Group, MESA+ Institute of Nanotechnology
	University of Twente, Enschede, Netherlands}
	\affiliation{China Academy of Space Technology, Xi'an, China}
\author{Kaixuan Ye}
\affiliation{Nonlinear Nanophotonics Group, MESA+ Institute of Nanotechnology
	University of Twente, Enschede, Netherlands}
\author{Roel Botter}
\affiliation{Nonlinear Nanophotonics Group, MESA+ Institute of Nanotechnology
	University of Twente, Enschede, Netherlands}
\author{Yvan Klaver}
\affiliation{Nonlinear Nanophotonics Group, MESA+ Institute of Nanotechnology
	University of Twente, Enschede, Netherlands}
\author{Qinggui~Tan}
\affiliation{China Academy of Space Technology, Xi'an, China}
\author{Hongxi~Yu}
\affiliation{China Academy of Space Technology, Xi'an, China}
\author{Marcel~Hoekman}
\affiliation{LioniX International BV, Enschede, Netherlands}
\author{Edwin~Klein}
\affiliation{LioniX International BV, Enschede, Netherlands}
\author{Chris Roeloffzen}
\affiliation{LioniX International BV, Enschede, Netherlands}
\author{Yang Liu}
\affiliation{Institute of Physics, Swiss Federal Institute of Technology Lausanne (EPFL), Lausanne, Switzerland}
\author{David Marpaung}
\email{david.marpaung@utwente.nl}
\affiliation{Nonlinear Nanophotonics Group, MESA+ Institute of Nanotechnology
	University of Twente, Enschede, Netherlands}

\date{\today}

\begin{abstract}
Microwave photonics (MWP) has adopted a number of important concepts and technologies over the recent pasts, including photonic integration, versatile programmability, and techniques for enhancing key radio frequency performance metrics such as the noise figure and the dynamic range. However, to date, these aspects have not been achieved simultaneously in a single circuit. Here, we demonstrate, for the first time, a multi-functional integrated microwave photonic circuit that enables on-chip programmable filtering functions while achieving record-high key radio frequency metrics of $>$120 dB.Hz dynamic range and 15 dB of noise figure that are previously unreachable. We unlock this unique feature by versatile complex spectrum tailoring using an all integrated modulation transformer and a double injection ring resonator as a multi-function optical filtering component. This work breaks the conventional and fragmented approach of integration, functionality and performance that currently prevents the adoption of integrated MWP systems in real applications.  

\end{abstract}

\maketitle

\section*{\label{sec:one}Introduction}

Integrated microwave photonic (MWP) systems can offer significant advantages in future radio frequency (RF) and microwave systems to realize advanced concepts such as multi-band, all-spectrum communications \cite{Akyildiz2020} and broadband programmable front-ends \cite{Chappell2014}, which are important for next generation communications (for example 6G). To play a key role in real RF applications, integrated MWP circuits need to simultaneously show advanced programmability and exceptional performance in terms of losses, noise figure, and dynamic range in a reduced footprint \cite{capmany2007microwave,yao2009microwave,capmany2012microwave,marpaung2013integrated,marpaung2019integrated}. In recent pasts, a number of integrated MWP functions have widely been  demonstrated, such as for filtering \cite{marpaung2013si,shahnia2015independent,fandino2017monolithic,liu2020integrated,tao2021hybrid}, phase shifting \cite{li2011broadband,burla2014onchip,porzi2018photonic,chew2019integrated,mckay2019brillouin}, delay line \cite{zhuang2011low,liu2017gigahertz,ji2019onchip,tsokos2021silicon}, waveform generator \cite{khan2010ultrabroad,weiner2011ultrafast,marpaung2011impulse,wang2015reconfigurable,zhu2020si,falconi2022} and beamforming \cite{yi2011photonic,yi2011photonic,zhuang2014chip,miller2020large,zhu2020silicon}. Typically, these functions were achieved in dedicated, or application specific circuits, and the measured RF metrics were only sparsely reported. The values of the RF gain, noise figure (NF), and spurious-free dynamic range (SFDR) in these circuits are usually far-off from the requirements of practical RF systems. 

Recently, new ways of building a programmable integrated MWP circuit have been extensively explored, for example through mesh-waveguide circuits that can synthesize a large-number of functionalities through programming \cite{zhuang2015programmable,perez2017multipurpose,perez2020multipurpose,bogaerts2020}. While versatile, the typical functional performance of these circuits, such as filter extinction, or passband quality, and the RF metrics were significantly lower than that of application specific circuits \cite{liu2017all,liu2018link,zhu2019positive,daulay2021microwave}. Thus, at present, the unique and well-sought combination of high integration-density, versatile programmability, and high RF performance remains elusive. In particular, there is a bottleneck in co-integrating components that support NF reduction and linearization with components that provide functionalities. 

Achieving simultaneous on-chip programmability and high RF performance requires integration of new types of photonic devices. A versatile modulation transformer (MT) that can tailor the phases and amplitudes of optical carrier and RF sidebands has recently identified as a key element for enhancement of dynamic range \cite{guo2021versatile,daulay2021tutorial,liu2021integrated}. On the other hand, a programmable resonator such as a double-injected ring resonator (DI-RR) \cite{cohen2018response} can provide a large number of filtering functions in a device. 

\begin{figure*}
\centering
\includegraphics[width=0.85\textwidth]{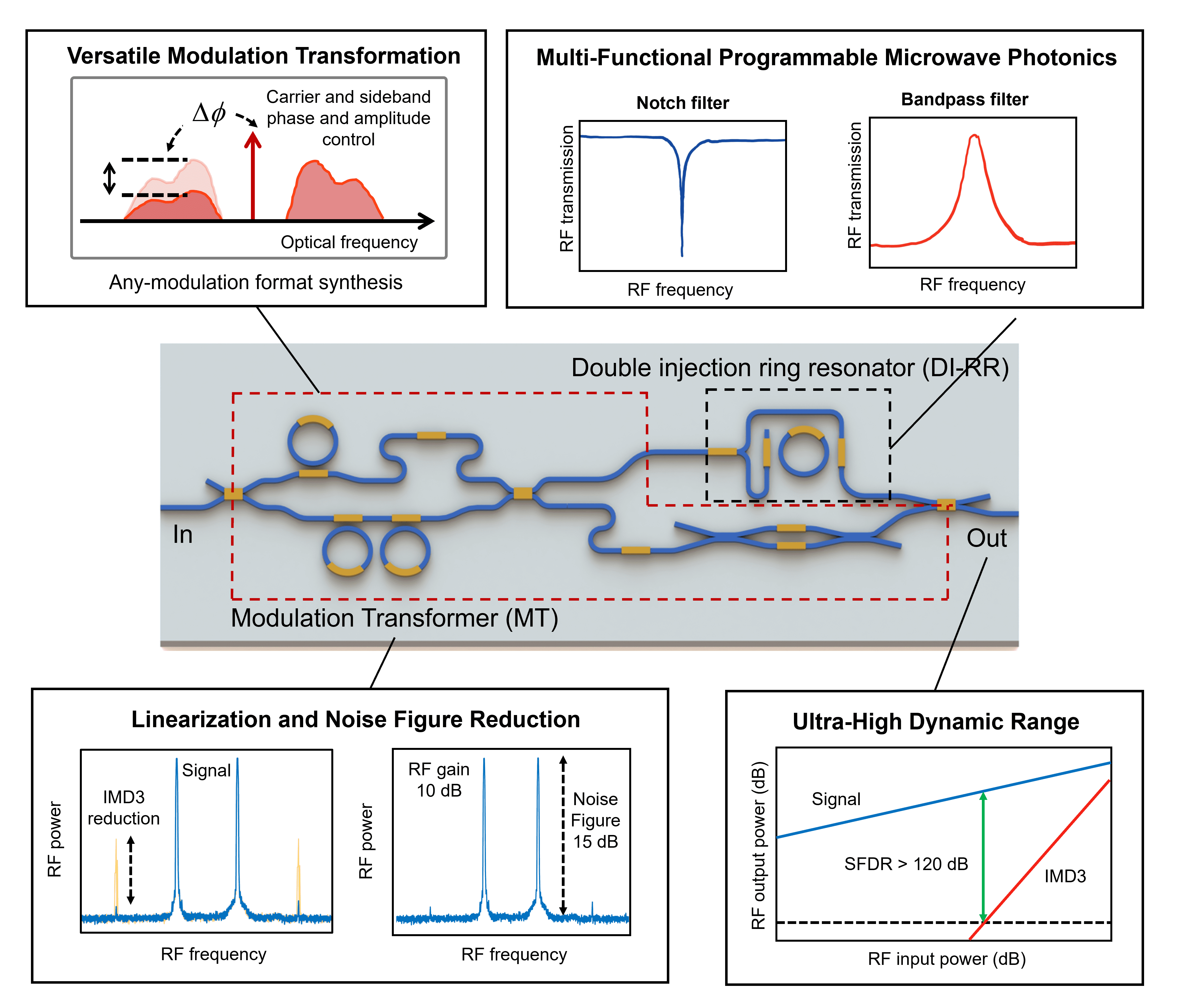}
\caption{\textbf{Artistic impression of an ultra-high dynamic range programmable integrated MWP circuit.} The  circuit contains of a versatile modulation transformer (MT) to independently tailor the phase and amplitude of optical modulation spectrum and an equally versatile double-injection ring resonator (DI-RR) to synthesize a variety of responses, including programmable RF filters. The combination of MT and DI-RR also allows for linearization through cancellation of intermodulation distortion (IMD) and noise figure (NF) reduction through low biasing and carrier suppression technique, leading to ultra-high dynamic range. SFDR: spurious-free dynamic range.}
\label{fig:fig1}
\vspace{-0.1cm}
\end{figure*}

In this work, we experimentally demonstrate a programmable integrated MWP circuit with a unique combination of a versatile MT device and a DI-RR, realized in a low-loss silicon nitride platform. With this circuit, we show for the first time, an array of RF filtering functions simultaneously with record-low NF of 15 dB, achieved using low-biasing technique, and ultra-high dynamic range of $>$ 120 dB.Hz, achieved using on-chip linearization. Our results, point a significant step forward towards the realization of practical programmable integrated MWP circuit in a real life applications.

\begin{figure*}
\centering
\includegraphics[width=\textwidth]{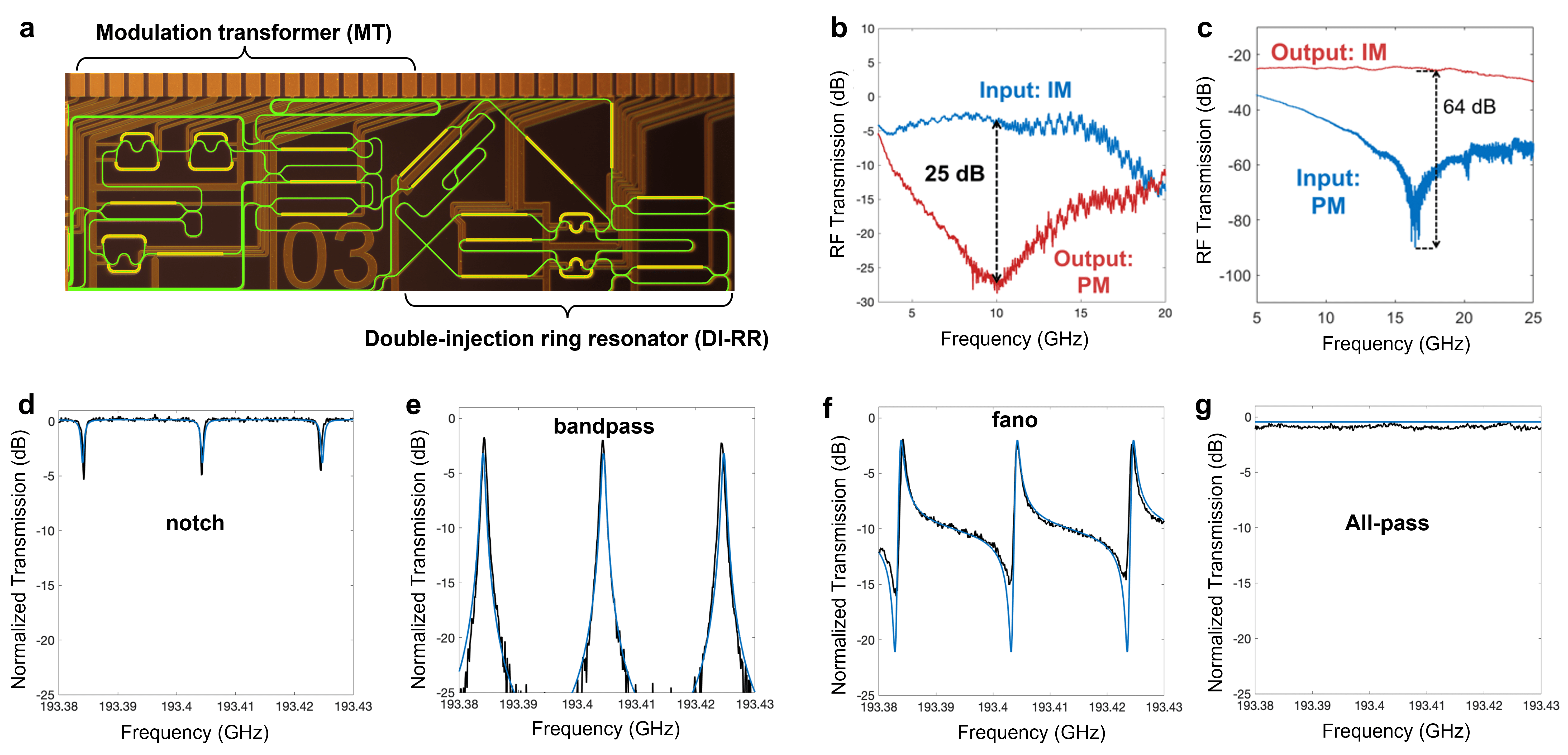}
\caption{\textbf{Programmable basic building blocks.} \textbf{(a)} Picture of the photonic chip, containing of a modulation transformer (MT) and a double injection ring resonator (DI-RR). The MT consists of a spectral de-interleaver used to isolate one RF sideband from the entire modulated spectrum. The phase shifter and tunable coupler are used to control the phase and amplitude of the isolated sideband. Modulation transformation is achieved after recombination at the output. See Supp. Info. A for details of the MT. \textbf{(b)} Experimental result of intensity-to-phase modulation (IM-PM) conversion with 25~dB extinction, achieved using the MT with IM input. \textbf{(c)} Measured phase-to-intensity modulation (PM-IM) conversion with 64~dB extinction, achieved using the MT with PM input. The DI-RR circuit is used to synthesize multiple filter functionalities. \textbf{(d-g)} Simulated (blue line) and measured (black line) selected responses of the silicon nitride DI-RR when tuned to exhibit notch filter \textbf{(d)}, bandpass filter \textbf{(e)},  Fano-like response \textbf{(f)}, and all-pass response \textbf{(g)}. See Supp. Info. B for details of the DI-RR.}
\label{fig:fig2}
\vspace{-0.1cm}
\end{figure*}

\vspace{-0.3cm}
\section*{Results}
\label{sec:two}
\subsection*{Programmable Microwave Photonics}
\label{susec:one}

The concept of ultrahigh dynamic range and low NF programmable integrated MWP filter explored here is illustrated in Fig.~\ref{fig:fig1}. The MT shapes the input modulated signal from an optical modulator to a desired modulation format that suited the targeted filtering function. Further, the spectral shaping can be used to create bespoke optical carrier and sidebands phase and amplitude distribution leading to linearization and NF reduction. The DI-RR is used as a filtering element that can provide notch or bandpass filtering from the same output port. In this way, high dynamic range and programmable filtering can be achieved simultaneously. 

The MT is implemented as an asymmetric Mach-Zehnder interferometer (aMZI) loaded with 3 ring resonators as a spectral de-interleaver \cite{luo2010high, guo2021versatile}, combined with a tunable attenuator and a phase shifter (For the details of the MT, see Supplementary information A). The  DI-RR is an add-drop ring resonator that is being double-injected from single input, hence capable of synthesizing a large number of responses (see Supplementary information B for the details of the DI-RR). The schematic of the entire circuit overlaid on a photograph of the fabricated silicon nitride chip is shown in Fig.~\ref{fig:fig2}a. The waveguide propagation loss in the circuit is 0.1 dB/cm, and the fiber-to-chip coupling loss is 1.1 dB/facet. 

The programmability of our circuit is demonstrated by measurement results depicted in Fig.~\ref{fig:fig2}b-\ref{fig:fig2}g. An optical modulated spectrum at the input of the MT circuit can be transformed into an output spectrum consisting of spectral components with designer phase and amplitude profiles. As examples, when operated with intensity modulator input, our silicon nitride MT circuit is capable of intensity-to-phase modulation (IM-PM) conversion with 25 dB extinction (Fig.~\ref{fig:fig2}b), and with a phase modulator input, the circuit can achieve PM-IM conversion with $>$ 60 dB of extinction (Fig.~\ref{fig:fig2}c). The DI-RR, on the other hand, can synthesize a variety of responses from a single output port, including notch filter, bandpass filter, Fano-like response, and an all-pass response \cite{okky2021mwp} as shown in Fig.~\ref{fig:fig2}d-\ref{fig:fig2}g. Because of the low propagation loss of the silicon nitride waveguide, narrow linewidth of the filtering function of around 400 MHz can be achieved in a device with 20 GHz free spectral range (FSR).   

\begin{figure*}
\centering
\includegraphics[width=\textwidth]{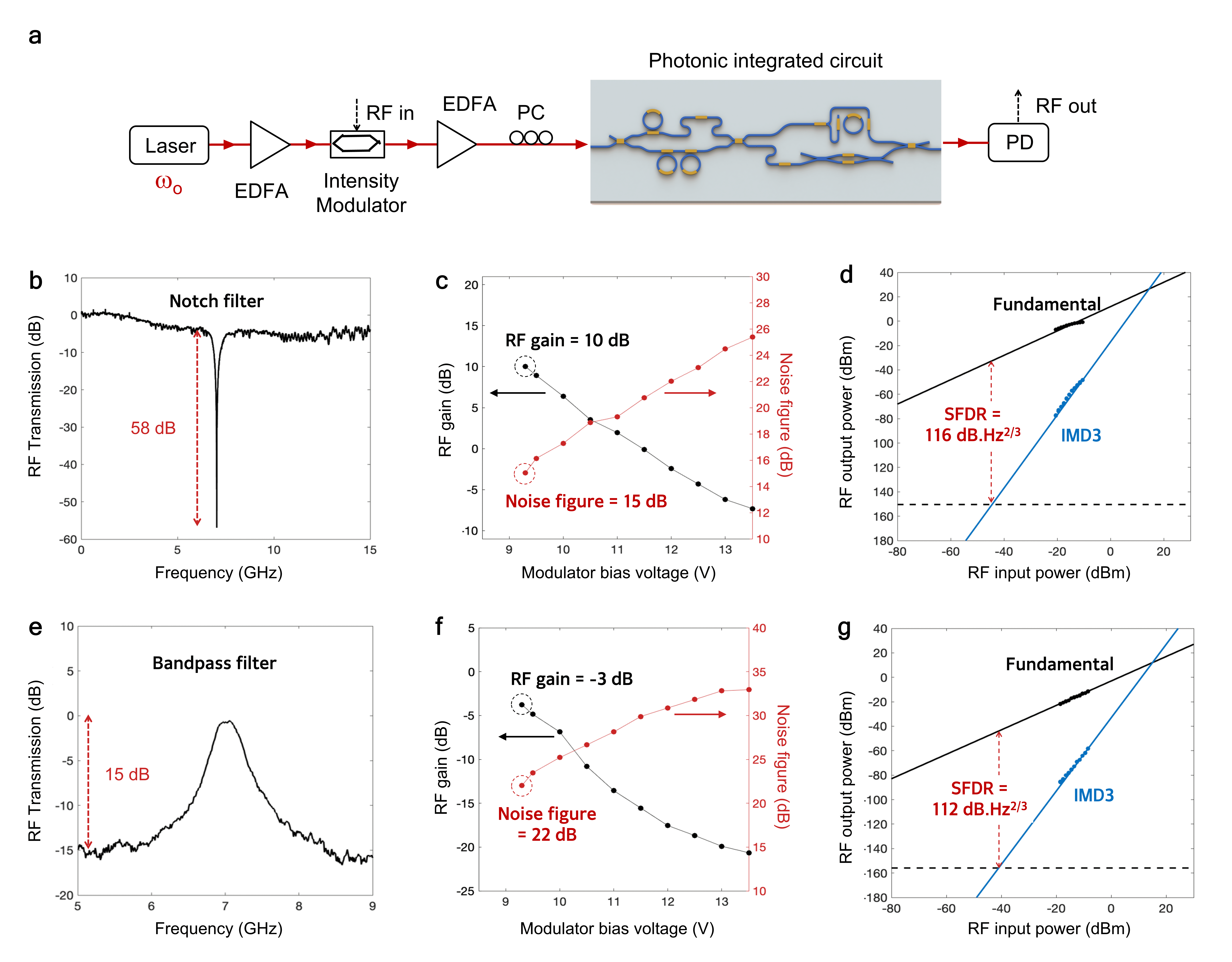}
\caption{\small \textbf{Noise figure reduction of the programmable MWP filters.} \textbf{a} Experiment setup of programmable microwave photonics with enhanced performance using low-biasing Mach-Zehnder Modulator (MZM). \textbf{b} High-rejection RF notch filter with 58 dB rejection. \textbf{c}, The measured RF gain and noise figure (NF) of the RF notch filter. Maximum RF gain of 10~dB and a minimum NF of 15~dB are achieved through the low-biasing of the MZM. \textbf{d}, The measured spurious-free dynamic range (SFDR) of the RF notch filter at 1 GHz, reaching 116 ~$\rm{dB}\cdot \rm{Hz}^{2/3}$. \textbf{e}, Single-sideband (SSB) RF bandpass filter with 15 dB rejection, obtain using optical carrier re-insertion with the MT and a bandpass response from the DI-RR. \textbf{f}, The measured RF gain and NF of the RF bandpass filter. Maximum RF gain of -3 dB and minimum noise figure of 22 dB are achieved using the low-biasing MZM. \textbf{g}, The measured SFDR  of the RF bandpass filter at 7 GHz, reaching a high value of 112~$\rm{dB}\cdot \rm{Hz}^{2/3}$. EDFA: erbium-doped fiber amplifier, MZM: Mach-Zehnder modulator, PC: polarization controller, RF: radio frequency, PD: photodetector}
\label{fig:fig3}
\vspace{-0.1cm}
\end{figure*}

\begin{figure*}[t]
\centering
\includegraphics[width=\textwidth]{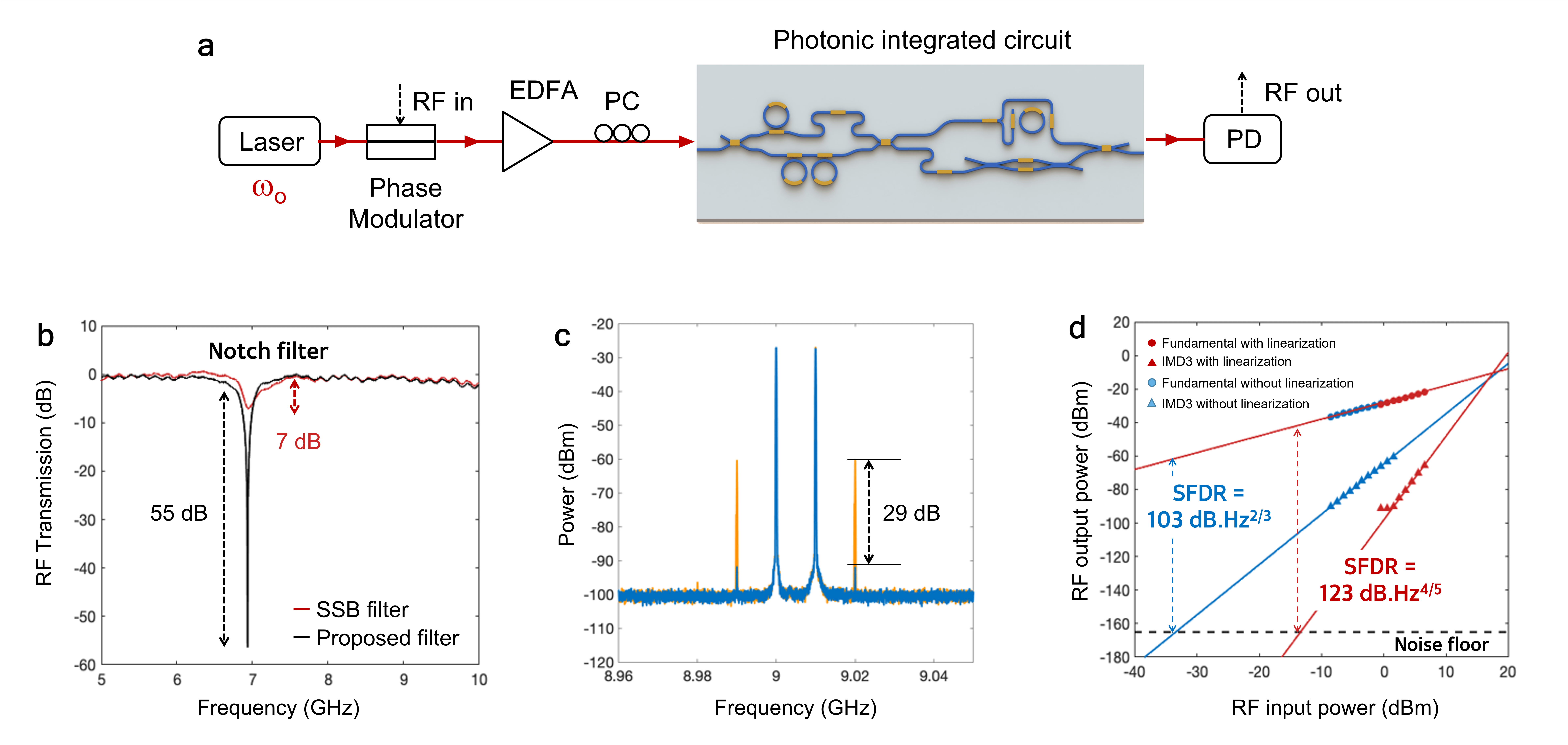}
\caption{\textbf{Linearization of RF notch filter.} \textbf{a} Experiment setup of simultaneous RF notch filter with linearization in a phase modulator (PM)-based system. \textbf{b}, Measured RF notch filter with 55 dB rejection. \textbf{c}, The measured two-tone RF spectra at the output of the photodetector for the linearized RF notch filter (blue) and single-sideband (SSB) RF notch filter without linearization (yellow) with 29 dB reduction of IMD3 power was achieved. \textbf{d}, The measured spurious-free dynamic range (SFDR) of the proposed linearized RF notch filter and SSB RF notch filter without linearization at RF frequency of 9~GHz. The proposed linearized RF notch filter has a record-high SFDR of 123 ~$\rm{dB}\cdot \rm{Hz}^{4/5}$. EDFA: erbium-doped fiber amplifier, PM: Phase modulator, PC: polarization controller, RF: radio frequency, PD: photodetector}
\label{fig:fig4}
\vspace{-0.1cm}
\end{figure*}

\vspace{-0.3cm}
\subsection*{High Gain and Low Noise Figure MWP Filter}
\label{susec:two}

We carried out a series of MWP filtering experiments using our circuit, with a setup diagrammed in Fig.~\ref{fig:fig3}a. A high power optical source implemented with a external cavity laser and an erbium-doped fiber amplifier (EDFA) is sent to a low-biased IM, creating an intensity modulated spectrum. We then configure our MT to shape the IM spectrum into a dedicated output spectrum suitable for the filtering tasks provided by the DI-RR (see Supplementary information C for additional details of the extended experimental scenarios).  In particular, we synthesized two RF filters with simultaneously high RF performance metrics. The first filter is an RF notch filter with intensity-to-asymmetric dual sidebands (IM-aDSB) modulation transformation using the MT and a notch response from the DI-RR. The second filter is an RF bandpass filter using combination of optical carrier re-insertion technique using the MT and a bandpass response from the DI-RR.

Figure~\ref{fig:fig3}b shows the measured RF notch filter with 58 dB rejection, 3-dB bandwidth of 400 MHz at the central frequency of 7 GHz. The high rejection in the filter was achieved through phase cancellation \cite{marpaung2015low} enabled by adjusting the phase and amplitude of the isolated sideband in IM-aDSB modulation conversion. Through low biasing the IM, we then achieved an optimum measured RF gain of 10 dB and a lowest measured NF of 15 dB (Fig.~\ref{fig:fig3}c). These high gain and low noise figure translated into a high spurious-free dynamic range (SFDR) of 116 ~$\rm{dB}\cdot \rm{Hz}^{2/3}$ when we performed a two-tone test at the frequency of 1 GHz (Fig.~\ref{fig:fig3}d). These results represent the best RF gain, noise figure, and SFDR combination for an MWP notch filter. The RF notch filter central frequency can be tuned over 20 GHz, which is the full FSR of the DI-RR. 

We then program the MT and the DI-RR to exhibit RF bandpass filter with 15 dB rejection, as shown in Fig.~\ref{fig:fig3}e. We measured RF gain of -3 dB, NF of 22 dB (Fig.~\ref{fig:fig3}f), and SFDR of 112 ~$\rm{dB}\cdot \rm{Hz}^{2/3}$ at 7 GHz (Fig.~\ref{fig:fig3}g). The versatility of the MT allows for the combination of low biasing and single-sideband (SSB)-based RF bandpass filtering, which would have not been possible only with an IM or a dual-parallel MZM. As a result, we can demonstrate the best combination of measured NF and SFDR for an RF bandpass filter. Importantly, these results were achieved without any electrical amplification in the system.

\begin{table*}[t]
\caption{Performance Comparison of Programmable Microwave Photonic Circuits}
\label{tab:res}
\vspace{0.3cm}
\begin{tabular}{l|c|c|c|c|c|c|c|c|c}
   \textbf{Year} &\textbf{Technology} & \textbf{Type of} & \textbf{Number of} & {\textbf{Type of}} & {\textbf{Tuning}} & {\textbf{Performance}}& {\textbf{Gain}} & {\textbf{Noise}} & {\textbf{SFDR}}\\
    & \textbf{platform} & \textbf{devices} & \textbf{functions} &\textbf{functions}  & \textbf{range} & \textbf{enhancement} & \textbf{} & \textbf{figure} & \textbf{}\\
     & &  &  &  & (GHz) &  & (dB) & (dB) & ($ \rm{dB}\cdot \rm{Hz}^{2/3}$)\\
    \hline
    2017 \cite{fandino2017monolithic} & InP & Las., Mod., & 1 & LPF & 0-6 & No & -20 & N/A & 81.4\\
     &  & RR, PD &  &  &  &  & &  & \\
    2018 \cite{perez2017multipurpose} & SOI & MZI mesh & 20 & BPF, notch & N/A & No & N/A & N/A & N/A\\
    2017 \cite{liu2017all} & Si$_3$N$_4$ & RR & 1 & Notch & 0-12 & NF:LB & 8 & 15.6 & 116\\
    2018 \cite{Zhang18OL} & SOI & Mod., RR, PD & 1 & BPF & 3-10 & No & -39 & N/A & 92.4\\
    2019  \cite{liu2019integrationbrillouin} & As$_2$S$_3$ & RR, SBS & 1 & Notch & 0-15 & NF:LB & -10 & 27.1 & 96.5\\
    2019 \cite{Zou19LPR} & InP & Las., Mod., & 7 & BPF, notch & 8-15 & No & N/A & N/A & N/A\\
     &  & MMI &  & IFM, RFG &  &  &  & & \\
    2020 \cite{Gertler20APL} & SOI & SBS & 1 & BPF & 4-10 & No & -17 & 56.7 & 90.3\\
    2020 \cite{daulay2020chip} & Si$_3$N$_4$ & RR & 1 & BPF & 2-7 & NF:CS & -10 & 27 & N/A\\
    2021 \cite{daulay2021microwave} & Si$_3$N$_4$ & RR & 1 & Notch & 3-10 & NF:CS & 3 & 31 & 100\\
    2021 \cite{guo2021versatile} & SOI & Mod., MT, & 2 & Notch, BPF & 5-25 & No & N/A & N/A & N/A\\
     &  & RR, PD &  &  &  &  & &  & \\
    2021 \cite{tao2021hybrid} & InP+SOI & Las., Mod., & 2 & Notch, BPF & 3-25 & No & -28 & 51 & 99.7\\
     &  & RR, PD &  &  &  &  & &  & \\
    2021 \cite{garret2021ultradeep} & Si$_3$N$_4$+As$_2$S$_3$ & RR, SBS & 1 & Notch & 2-12 & No & N/A & N/A & 92.2\\
    2021 \cite{Li2021OpeX} & Si$_3$N$_4$+LiNbO$_3$ & Mod., RR & 1 & Downconversion & 4-20 & No & -10 & 45 & 105\\
    \textbf{(this work)} & Si$_3$N$_4$  & MT, DI-RR & 6 & Notch, BPF & 4-20 & NF:LB & \textbf{10} & \textbf{15} & 116\\
    \textbf{(this work)} & Si$_3$N$_4$  & MT, DI-RR & 6 & Notch & 6-18 & SFDR:Lin. & -26 & 35 & \textbf{123}
\vspace{0.3cm}
\end{tabular}

{\justifying RR: ring resonator, PD: photodetector, Mod: modulator, MMI: multi-mode interference, SBS: stimulated Brillouin scattering, MT: modulation transformer, DI-RR: double-injection ring resonator, LPF: low pass filter, BPF: bandpass filter, IFM: instantaneous frequency measurement,  PS: phase shifter, NF: noise figure, LB: low biasing, CS: carrier suppression, SFDR: spurious-free dynamic range, Lin: linearization.\par}
\vspace{-0.1cm}
\end{table*}

\vspace{-0.3cm}
\subsection*{On-Chip Linearization for Ultrahigh Dynamic Range}
\label{susec:three}

Some demanding applications of MWP require ultra-high SFDR beyond 120 dB.Hz \cite{marpaung2019integrated}. Such a metric can be achieved through linearization of the modulator transfer function \cite{ZhangOoptica2016,ZhangOL20,MortonOE21,feng2022ultrahighlinearity}. But achieving linearization simultaneously with functionalities has been proven challenging. Exploiting the versatility of our MT and DI-RR circuit, we demonstrate, for the first time, simultaneous RF notch filter and linearization in the same photonic chip.

Here, we implement third-order intermodulation distortion (IMD3) cancellation technique using complex multi-order sidebands spectral shaping as we previously reported \cite{liu2020linearized,liu2021integrated,liu2021simultaneous} (see Supplmentary Notes D and E for the details of the linearization technique). In this experiment, light is sent to a phase modulator (PM) (Fig.~\ref{fig:fig4}a) that generates the optical carrier and first and higher order sidebands. The MT is used to convert the PM to aDSB modulation format, and to shape the amplitudes and phases of the first and second order sidebands yielding to IMD3 suppression. The DI-RR is then used to filter the first order sideband to create a phase cancellation RF notch filter. In this way,  simultaneous RF notch filter and linearization can be achieved, leading to significant enhancement of the  SFDR (For additional details of linearization theoretical analysis and working principle, see Supplementary information D and E respectively). 

A conventional SSB RF notch filter without any linearization was used as a benchmark for the proposed linearized RF notch filter. Figure ~\ref{fig:fig4}b shows the proposed linearized RF notch filter with $>$ 45 dB rejection improvement compared to the conventional SSB RF notch filter. With the linearization, we achieved 29 dB of IMD3 suppresion (Fig.~\ref{fig:fig4}c) and $>$ 20 dB SFDR improvement (Fig.~\ref{fig:fig4}d). With a noise floor of -164 dBm/Hz, the conventional SSB RF notch filter shows an SFDR of 103 ~$\rm{dB}\cdot \rm{Hz}^{2/3}$, while the linearized RF notch filter exhibits a record-high SFDR of 123 ~$\rm{dB}\cdot \rm{Hz}^{4/5}$. It is clear from the slope of the measured IMD3 powers that the fifth order distortion becomes the dominant factor due to cancellation of the third-order distortion. This result marked the highest SFDR simultaneously achieved with on-chip signal functionality.

\vspace{-0.3cm}
\section*{Discussion}
\label{sec:three}
Table I summarizes the performance comparison of existing programmable MWP circuits reported in the past 5 years. The route to largest number of functions is still provided through MZI-mesh circuits \cite{perez2017multipurpose}. Virtually all demonstrations, except the application specific RF notch filter reported in \cite{liu2017all}, show relatively low RF performance, notably characterised by very high NF ($>$25 dB) and low SFDR ($<$ 100 dB.Hz). These results highlight the challenge in creating multifunctional and high-performance MWP circuit. Our results overcome this challenge by delivering high-level of reconfigurability with variety of functions in combination with  high gain, low NF, and ultrahigh SFDR. Importantly, in the context of functional circuits, our demonstrations represent record-low NF and record-high SFDR to date.

At present, our approach enables NF reduction through low biasing of MZM intensity modulator. This translates into moderate SFDR enhancement as seen in Table I. Our linearization method, on the other hand, is working for PM-based MWP filter where maximum IMD3 suppression is achieved when the optical carrier is partially suppressed (See Supplementary Info D). Due to the limitation in maximum optical amplification in our experimental setup, the measured ultra-high SFDR is still accompanied by relatively low link gain and high NF (Table 1). Achieving the gain, NF, and SFDR advantages simultaneously is feasible in the PM link when higher optical amplification and higher power handling components are used. Alternatively. this can also be achieved by developing similar linearization strategy for low-biased MZM link.

In summary, we have designed, fabricated, and experimentally demonstrated, for the first time, programmable integrated MWP circuit based on a unique combination of versatile MT and equally versatile DI-RR. We reconfigure the circuit to synthesize an array of RF filtering functions with high RF gain, low NF, and ultrahigh SFDR concurrently. Hence, this work open a new paradigm and an important step forward towards the realization of programmable integrated MWP circuit with versatile functions, low NF, and ultra-high dynamic range for real life applications.

\vspace{-0.3cm}
\section*{Methods}
\label{sec:four}
\subsection*{Silicon Nitride Circuit Fabrication}

The waveguides in our circuit are fabricated using LioniX standard TriPleX Asymmetric Double-Stripe (ADS) geometry \cite{worhoff2015triplex,roeloffzen2018low}. First, a $\rm SiO_2$ layer is grown from wet thermal oxidation of single-crystal silicon substrate with temperatures equal to or above 1000 \textcelsius. Then, low-pressure chemical vapor deposition (LPCVD) is used for the Si$_3$N$_4$ layers and together with the gas tetraethylorthosilicate (TEOS) for intermediate SiO$_2$ layer. Next, the waveguides are patterned using contact lithography, and processed with dry etching. Last, the waveguides are covered with an additional SiO$_2$ layer through LPCVD TEOS. Because typical top cladding thickness cannot achieved only by LPCVD TEOS, plasma-enhanced chemical vapor deposition (PECVD) is used to increase the SiO$_2$ top cladding thickness to a total of 8~µm. A thicker layers can be achieved because of the stress in PECVD SiO$_2$ layers is much less than LPCVD layers.

\subsection*{Details of Programmable Microwave Photonics  Experiments}

An optical carrier from low relative-intensity noise (RIN) laser (Pure Photonics PPCL550) with RIN of -155 dB/Hz and wavelength set at 1550 nm is amplified with a low-noise erbium-doped fiber amplifier (EDFA, Amonics). Then, the output of amplified optical carrier is optically modulated using a MZM (Thorlabs, LN05S-FC 40 GHz) with bias point set at quadrature ($\theta_{\rm{B}}$ = $\pi$/2). The MZM is driven by RF signal from a vector network analyzer (VNA, Keysight P5007A). The output of the MZM then sent to another low-noise EDFA (Amonics) before being injected into a programmable silicon nitride chip (LioniX International BV) fabricated using low-loss TriPleX (Si$_3$N$_4$/SiO$_2$) technology\cite{worhoff2015triplex,roeloffzen2018low} with propagation loss of the optical waveguide at 0.1~dB/cm. The chip can be tuned through thermo-optic tuning using custom made heater controller software, while being stabilized by a thermoelectric cooler (TEC) controller. The processed optical signal is sent to a photodetector (PD, APIC 40 GHz) and the converted RF signal is measured with a VNA, while the RF spectrum analyzer (RFSA, Keysight N9000B) is used to measure the RF filter's noise and linearity.

\subsection*{Details of Linearization Experiments}

A low RIN laser (Pure Photonics PPCL550) is modulated using a PM (EOSpace 20 GHz) using a sweeping RF signal with -3~dBm power from VNA (Keysight P5007A) for RF notch filter experiment. For the dynamic range experiment, the two-tone RF signal with power of 10~dBm, centered at 9~GHz with a space of 10~MHz from signal generators (Wiltron 69147A and Rohde-Schwarz SMP02) is used to drive the PM. The modulated signal is then amplified by a low-noise EDFA (Amonics), before coupling to a programmable silicon nitride chip (LioniX International BV). The processed optical spectrum is sent to a PD (APIC 40 GHz) to retrieve the RF signal and measured using the VNA for the filter's response and the RFSA (Keysight N9000B) for linearity.

\section*{Author Contributions}
O.D. and G.L. contributed equally in this work. O.D., G.L.,and D.M. developed the concept and proposed the physical system.  O.D. designed the photonic circuits, O.D. and G.L. developed and performed numerical simulations. O.D. and G.L. performed the experiments with input from K.Y., R.B., and Y.K. E.K., M.H., and C.R. layout, developed, and fabricated the silicon nitride circuits.  D.M., O.D. and G.L. wrote the manuscript with input from all authors. D.M. led and supervised the entire project.

\section*{ Funding Information}
This work was supported by Netherlands Organisation for Scientific Research NWO Vidi (15702) and NWO Start Up (740.018.021).

\bibliographystyle{IEEEtran} 
\bibliography{library}

\begin{thebibliography}{10}
\providecommand{\url}[1]{#1}
\csname url@samestyle\endcsname
\providecommand{\newblock}{\relax}
\providecommand{\bibinfo}[2]{#2}
\providecommand{\BIBentrySTDinterwordspacing}{\spaceskip=0pt\relax}
\providecommand{\BIBentryALTinterwordstretchfactor}{4}
\providecommand{\BIBentryALTinterwordspacing}{\spaceskip=\fontdimen2\font plus
\BIBentryALTinterwordstretchfactor\fontdimen3\font minus
  \fontdimen4\font\relax}
\providecommand{\BIBforeignlanguage}[2]{{%
\expandafter\ifx\csname l@#1\endcsname\relax
\typeout{** WARNING: IEEEtran.bst: No hyphenation pattern has been}%
\typeout{** loaded for the language `#1'. Using the pattern for}%
\typeout{** the default language instead.}%
\else
\language=\csname l@#1\endcsname
\fi
#2}}
\providecommand{\BIBdecl}{\relax}
\BIBdecl

\bibitem{Akyildiz2020}
I.~F. Akyildiz, A.~Kak, and S.~Nie, ``6g and beyond: The future of wireless
  communications systems,'' \emph{IEEE Access}, vol.~8, pp. 133\,995--134\,030,
  2020.

\bibitem{Chappell2014}
W.~J. Chappell, E.~J. Naglich, C.~Maxey, and A.~C. Guyette, ``Putting the radio
  in “software-defined radio”: Hardware developments for adaptable rf
  systems,'' \emph{Proceedings of the IEEE}, vol. 102, no.~3, pp. 307--320,
  2014.

\bibitem{capmany2007microwave}
J.~Capmany and D.~Novak, ``Microwave photonics combines two worlds,''
  \emph{Nature photonics}, vol.~1, no.~6, p. 319, 2007.

\bibitem{yao2009microwave}
J.~Yao, ``Microwave photonics,'' \emph{Journal of lightwave technology},
  vol.~27, no.~3, pp. 314--335, 2009.

\bibitem{capmany2012microwave}
J.~Capmany, J.~Mora, I.~Gasulla \emph{et~al.}, ``Microwave photonic signal
  processing,'' \emph{Journal of Lightwave Technology}, vol.~31, no.~4, pp.
  571--586, 2012.

\bibitem{marpaung2013integrated}
D.~Marpaung, C.~Roeloffzen, R.~Heideman \emph{et~al.}, ``Integrated microwave
  photonics,'' \emph{Laser \& Photonics Reviews}, vol.~7, no.~4, pp. 506--538,
  2013.

\bibitem{marpaung2019integrated}
D.~Marpaung, J.~Yao, and J.~Capmany, ``Integrated microwave photonics,''
  \emph{Nature photonics}, vol.~13, no.~2, pp. 80--90, 2019.

\bibitem{marpaung2013si}
D.~Marpaung, B.~Morrison, R.~Pant \emph{et~al.}, ``Si 3 n 4 ring
  resonator-based microwave photonic notch filter with an ultrahigh peak
  rejection,'' \emph{Optics express}, vol.~21, no.~20, pp. 23\,286--23\,294,
  2013.

\bibitem{shahnia2015independent}
S.~Shahnia, M.~Pagani, B.~Morrison, B.~J. Eggleton, and D.~Marpaung,
  ``Independent manipulation of the phase and amplitude of optical sidebands in
  a highly-stable rf photonic filter,'' \emph{Optics express}, vol.~23, no.~18,
  pp. 23\,278--23\,286, 2015.

\bibitem{fandino2017monolithic}
J.~S. Fandi{\~n}o, P.~Mu{\~n}oz, D.~Dom{\'e}nech, and J.~Capmany, ``A
  monolithic integrated photonic microwave filter,'' \emph{Nature Photonics},
  vol.~11, no.~2, pp. 124--129, 2017.

\bibitem{liu2020integrated}
Y.~Liu, A.~Choudhary, D.~Marpaung, and B.~J. Eggleton, ``Integrated microwave
  photonic filters,'' \emph{Advances in Optics and Photonics}, vol.~12, no.~2,
  pp. 485--555, 2020.

\bibitem{tao2021hybrid}
Y.~Tao, H.~Shu, X.~Wang \emph{et~al.}, ``Hybrid-integrated high-performance
  microwave photonic filter with switchable response,'' \emph{Photonics
  Research}, vol.~9, no.~8, pp. 1569--1580, 2021.

\bibitem{li2011broadband}
W.~Li, N.~H. Zhu, L.~X. Wang, and H.~Wang, ``Broadband phase-to-intensity
  modulation conversion for microwave photonics processing using
  brillouin-assisted carrier phase shift,'' \emph{Journal of lightwave
  technology}, vol.~29, no.~24, pp. 3616--3621, 2011.

\bibitem{burla2014onchip}
M.~Burla, L.~R. Cortes, M.~Li \emph{et~al.}, ``On-chip programmable
  ultra-wideband microwave photonic phase shifter and true time delay unit,''
  \emph{Optics letters}, vol.~39, no.~21, pp. 6181--6184, 2014.

\bibitem{porzi2018photonic}
C.~Porzi, G.~Serafino, M.~Sans \emph{et~al.}, ``Photonic integrated microwave
  phase shifter up to the mm-wave band with fast response time in
  silicon-on-insulator technology,'' \emph{Journal of Lightwave Technology},
  vol.~36, no.~19, pp. 4494--4500, 2018.

\bibitem{chew2019integrated}
S.~X. Chew, D.~Huang, L.~Li \emph{et~al.}, ``Integrated microwave photonic
  phase shifter with full tunable phase shifting range ($>$ 360°) and rf power
  equalization,'' \emph{Optics express}, vol.~27, no.~10, pp. 14\,798--14\,808,
  2019.

\bibitem{mckay2019brillouin}
L.~McKay, M.~Merklein, A.~C. Bedoya \emph{et~al.}, ``Brillouin-based phase
  shifter in a silicon waveguide,'' \emph{Optica}, vol.~6, no.~7, pp. 907--913,
  2019.

\bibitem{zhuang2011low}
L.~Zhuang, D.~Marpaung, M.~Burla \emph{et~al.}, ``Low-loss, high-index-contrast
  si 3 n 4/sio 2 optical waveguides for optical delay lines in microwave
  photonics signal processing,'' \emph{Optics express}, vol.~19, no.~23, pp.
  23\,162--23\,170, 2011.

\bibitem{liu2017gigahertz}
Y.~Liu, A.~Choudhary, D.~Marpaung, and B.~J. Eggleton, ``Gigahertz optical
  tuning of an on-chip radio frequency photonic delay line,'' \emph{Optica},
  vol.~4, no.~4, pp. 418--423, 2017.

\bibitem{ji2019onchip}
X.~Ji, X.~Yao, Y.~Gan \emph{et~al.}, ``On-chip tunable photonic delay line,''
  \emph{APL Photonics}, vol.~4, no.~9, p. 090803, 2019.

\bibitem{tsokos2021silicon}
C.~Tsokos, E.~Andrianopoulos, A.~Raptakis \emph{et~al.}, ``True time delay
  optical beamforming network based on hybrid inp-silicon nitride
  integration,'' \emph{Journal of Lightwave Technology}, vol.~39, no.~18, pp.
  5845--5854, 2021.

\bibitem{khan2010ultrabroad}
M.~H. Khan, H.~Shen, Y.~Xuan \emph{et~al.}, ``Ultrabroad-bandwidth arbitrary
  radiofrequency waveform generation with a silicon photonic chip-based
  spectral shaper,'' \emph{Nature Photonics}, vol.~4, no.~2, pp. 117--122,
  2010.

\bibitem{weiner2011ultrafast}
A.~M. Weiner, ``Ultrafast optical pulse shaping: A tutorial review,''
  \emph{Optics Communications}, vol. 284, no.~15, pp. 3669--3692, 2011.

\bibitem{marpaung2011impulse}
D.~Marpaung, L.~Chevalier, M.~Burla, and C.~Roeloffzen, ``Impulse radio
  ultrawideband pulse shaper based on a programmable photonic chip frequency
  discriminator,'' \emph{Optics express}, vol.~19, no.~25, pp.
  24\,838--24\,848, 2011.

\bibitem{wang2015reconfigurable}
J.~Wang, H.~Shen, L.~Fan \emph{et~al.}, ``Reconfigurable radio-frequency
  arbitrary waveforms synthesized in a silicon photonic chip,'' \emph{Nature
  communications}, vol.~6, no.~1, pp. 1--8, 2015.

\bibitem{zhu2020si}
Z.~Zhu, Y.~Liu, M.~Merklein \emph{et~al.}, ``Si 3 n 4-chip-based versatile
  photonic rf waveform generator with a wide tuning range of repetition rate,''
  \emph{Optics letters}, vol.~45, no.~6, pp. 1370--1373, 2020.

\bibitem{falconi2022}
F.~Falconi, C.~Porzi, A.~Malacarne \emph{et~al.}, ``Uwb fastlytunable 0.550 ghz
  rf transmitter based on integrated photonics,'' \emph{Journal of Lightwave
  Technology}, pp. 1--1, 2021.

\bibitem{yi2011photonic}
X.~Yi, T.~X. Huang, and R.~A. Minasian, ``Photonic beamforming based on
  programmable phase shifters with amplitude and phase control,'' \emph{IEEE
  Photonics Technology Letters}, vol.~23, no.~18, pp. 1286--1288, 2011.

\bibitem{zhuang2014chip}
L.~Zhuang, M.~Hoekman, C.~Taddei \emph{et~al.}, ``On-chip microwave photonic
  beamformer circuits operating with phase modulation and direct detection,''
  \emph{Optics express}, vol.~22, no.~14, pp. 17\,079--17\,091, 2014.

\bibitem{miller2020large}
S.~A. Miller, Y.-C. Chang, C.~T. Phare \emph{et~al.}, ``Large-scale optical
  phased array using a low-power multi-pass silicon photonic platform,''
  \emph{Optica}, vol.~7, no.~1, pp. 3--6, 2020.

\bibitem{zhu2020silicon}
C.~Zhu, L.~Lu, W.~Shan \emph{et~al.}, ``Silicon integrated microwave photonic
  beamformer,'' \emph{Optica}, vol.~7, no.~9, pp. 1162--1170, 2020.

\bibitem{zhuang2015programmable}
L.~Zhuang, C.~G. Roeloffzen, M.~Hoekman, K.-J. Boller, and A.~J. Lowery,
  ``Programmable photonic signal processor chip for radiofrequency
  applications,'' \emph{Optica}, vol.~2, no.~10, pp. 854--859, 2015.

\bibitem{perez2017multipurpose}
D.~P{\'e}rez, I.~Gasulla, L.~Crudgington \emph{et~al.}, ``Multipurpose silicon
  photonics signal processor core,'' \emph{Nature communications}, vol.~8,
  no.~1, pp. 1--9, 2017.

\bibitem{perez2020multipurpose}
D.~P{\'e}rez, A.~Lopez, P.~DasMahapatra, and J.~Capmany, ``Multipurpose
  self-configuration of programmable photonic circuits,'' \emph{Nature
  communications}, vol.~11, no. 6359, pp. 1--11, 2020.

\bibitem{bogaerts2020}
W.~Bogaerts, D.~P{\'e}rez, J.~Capmany \emph{et~al.}, ``Programmable photonic
  circuits,'' \emph{Nature}, vol. 586, no. 7828, pp. 207--216, Oct 2020.

\bibitem{liu2017all}
Y.~Liu, J.~Hotten, A.~Choudhary, B.~J. Eggleton, and D.~Marpaung,
  ``All-optimized integrated rf photonic notch filter,'' \emph{Optics letters},
  vol.~42, no.~22, pp. 4631--4634, 2017.

\bibitem{liu2018link}
Y.~Liu, D.~Marpaung, A.~Choudhary, J.~Hotten, and B.~J. Eggleton, ``Link
  performance optimization of chip-based si 3 n 4 microwave photonic filters,''
  \emph{Journal of lightwave technology}, vol.~36, no.~19, pp. 4361--4370,
  2018.

\bibitem{zhu2019positive}
Z.~Zhu, Y.~Liu, M.~Merklein \emph{et~al.}, ``Positive link gain microwave
  photonic bandpass filter using si 3 n 4-ring-enabled sideband filtering and
  carrier suppression,'' \emph{Optics express}, vol.~27, no.~22, pp.
  31\,727--31\,740, 2019.

\bibitem{daulay2021microwave}
O.~Daulay, G.~Liu, and D.~Marpaung, ``Microwave photonic notch filter with
  integrated phase-to-intensity modulation transformation and optical carrier
  suppression,'' \emph{Optics letters}, vol.~46, no.~3, pp. 488--491, 2021.

\bibitem{guo2021versatile}
X.~Guo, Y.~Liu, T.~Yin \emph{et~al.}, ``Versatile silicon microwave photonic
  spectral shaper,'' \emph{APL Photonics}, vol.~6, no.~3, p. 036106, 2021.

\bibitem{daulay2021tutorial}
O.~Daulay, G.~Liu, X.~Guo, M.~Eijkel, and D.~Marpaung, ``A tutorial on
  integrated microwave photonic spectral shaping,'' \emph{Journal of Lightwave
  Technology}, vol.~39, no.~3, pp. 700--711, 2021.

\bibitem{liu2021integrated}
G.~Liu, O.~Daulay, Y.~Klaver \emph{et~al.}, ``Integrated microwave photonic
  spectral shaping for linearization and spurious-free dynamic range
  enhancement,'' \emph{Journal of Lightwave Technology}, 2021.

\bibitem{cohen2018response}
R.~Cohen, O.~Amrani, and S.~Ruschin, ``Response shaping with a silicon ring
  resonator via double injection,'' \emph{Nature Photonics}, vol.~12, no.~11,
  pp. 706--712, 2018.

\bibitem{luo2010high}
L.-W. Luo, S.~Ibrahim, A.~Nitkowski \emph{et~al.}, ``High bandwidth on-chip
  silicon photonic interleaver,'' \emph{Optics express}, vol.~18, no.~22, pp.
  23\,079--23\,087, 2010.

\bibitem{okky2021mwp}
O.~Daulay, G.~Liu, R.~Botter \emph{et~al.}, ``Reconfigurable double-injection
  ring resonator for integrated microwave photonic signal processing,'' in
  \emph{2021 International Topical Meeting on Microwave Photonics (MWP)}, Nov.
  2021, pp. 1--4.

\bibitem{marpaung2015low}
D.~Marpaung, B.~Morrison, M.~Pagani \emph{et~al.}, ``Low-power, chip-based
  stimulated brillouin scattering microwave photonic filter with ultrahigh
  selectivity,'' \emph{Optica}, vol.~2, no.~2, pp. 76--83, 2015.

\bibitem{Zhang18OL}
W.~Zhang and J.~Yao, ``On-chip silicon photonic integrated frequency-tunable
  bandpass microwave photonic filter,'' \emph{Opt. Lett.}, vol.~43, no.~15, pp.
  3622--3625, Aug 2018.

\bibitem{liu2019integrationbrillouin}
Y.~Liu, A.~Choudhary, G.~Ren \emph{et~al.}, ``Integration of brillouin and
  passive circuits for enhanced radio-frequency photonic filtering,'' \emph{APL
  Photonics}, vol.~4, no. 106103, pp. 106\,103--1--106\,103--12, 2019.

\bibitem{Zou19LPR}
X.~Zou, F.~Zou, Z.~Cao \emph{et~al.}, ``A multifunctional photonic integrated
  circuit for diverse microwave signal generation, transmission, and
  processing,'' \emph{Laser \& Photonics Reviews}, vol.~13, no.~6, p. 1800240,
  2019.

\bibitem{Gertler20APL}
S.~Gertler, E.~A. Kittlaus, N.~T. Otterstrom, and P.~T. Rakich, ``Tunable
  microwave-photonic filtering with high out-of-band rejection in silicon,''
  \emph{APL Photonics}, vol.~5, no.~9, p. 096103, 2020.

\bibitem{daulay2020chip}
O.~Daulay, R.~Botter, and D.~Marpaung, ``On-chip programmable microwave
  photonic filter with an integrated optical carrier processor,'' \emph{OSA
  Continuum}, vol.~3, no.~8, pp. 2166--2174, 2020.

\bibitem{garret2021ultradeep}
M.~Garret, Y.~Liu, D.-Y. Choi \emph{et~al.}, ``Ultra-deep multi-notch microwave
  photonic filter utilising on-chip brillouin processing and microring
  resonators,'' in \emph{2021 Conference on Laser and Electro-Optics (CLEO)
  Europe}, Jun. 2021, p.~1.

\bibitem{Li2021OpeX}
J.~Li, S.~Yang, H.~Chen, and M.~Chen, ``Hybrid microwave photonic receiver
  based on integrated tunable bandpass filters,'' \emph{Opt. Express}, vol.~29,
  no.~7, pp. 11\,084--11\,093, Mar 2021.

\bibitem{ZhangOoptica2016}
C.~Zhang, P.~A. Morton, J.~B. Khurgin, J.~D. Peters, and J.~E. Bowers,
  ``Ultralinear heterogeneously integrated ring-assisted mach\&\#x2013;zehnder
  interferometer modulator on silicon,'' \emph{Optica}, vol.~3, no.~12, pp.
  1483--1488, Dec 2016.

\bibitem{ZhangOL20}
Q.~Zhang, H.~Yu, P.~Xia \emph{et~al.}, ``High linearity silicon modulator
  capable of actively compensating input distortion,'' \emph{Opt. Lett.},
  vol.~45, no.~13, pp. 3785--3788, Jul 2020.

\bibitem{MortonOE21}
P.~A. Morton, J.~B. Khurgin, and M.~J. Morton, ``All-optical linearized
  mach-zehnder modulator,'' \emph{Opt. Express}, vol.~29, no.~23, pp.
  37\,302--37\,313, Nov 2021.

\bibitem{feng2022ultrahighlinearity}
H.~Feng, K.~Zhang, W.~Sun \emph{et~al.}, ``Ultra-high-linearity integrated
  lithium niobate electro-optic modulators,'' \emph{Arxiv}, pp. 1--16, 2022.

\bibitem{liu2020linearized}
G.~Liu, O.~Daulay, Q.~Tan, H.~Yu, and D.~Marpaung, ``Linearized phase modulated
  microwave photonic link based on integrated ring resonators,'' \emph{Optics
  express}, vol.~28, no.~26, pp. 38\,603--38\,615, 2020.

\bibitem{liu2021simultaneous}
G.~Liu, O.~Daulay, Q.~Tan \emph{et~al.}, ``Simultaneous notch filtering and
  linearization in an integrated microwave photonic circuit,'' in \emph{2021
  International Topical Meeting on Microwave Photonics (MWP)}, Nov. 2021, pp.
  1--4.

\bibitem{worhoff2015triplex}
K.~Worhoff, R.~G. Heideman, A.~Leinse, and M.~Hoekman, ``Triplex: a versatile
  dielectric photonic platform,'' \emph{Advanced Optical Technologies}, vol.~4,
  no.~2, pp. 189--207, 2015.

\bibitem{roeloffzen2018low}
C.~G. Roeloffzen, M.~Hoekman, E.~J. Klein \emph{et~al.}, ``Low-loss si3n4
  triplex optical waveguides: Technology and applications overview,''
  \emph{IEEE journal of selected topics in quantum electronics}, vol.~24,
  no.~4, pp. 1--21, 2018.

\bibitem{bogaerts2012silicon}
W.~Bogaerts, P.~De~Heyn, T.~Van~Vaerenbergh \emph{et~al.}, ``Silicon microring
  resonators,'' \emph{Laser \& Photonics Reviews}, vol.~6, no.~1, pp. 47--73,
  2012.

\end{thebibliography}

\newpage
\onecolumngrid
\beginsupplement
\newpage

\section*{Supplementary Information A: Modulation Transformer}

\begin{figure}[ht]
\centering
\includegraphics[width=\textwidth]{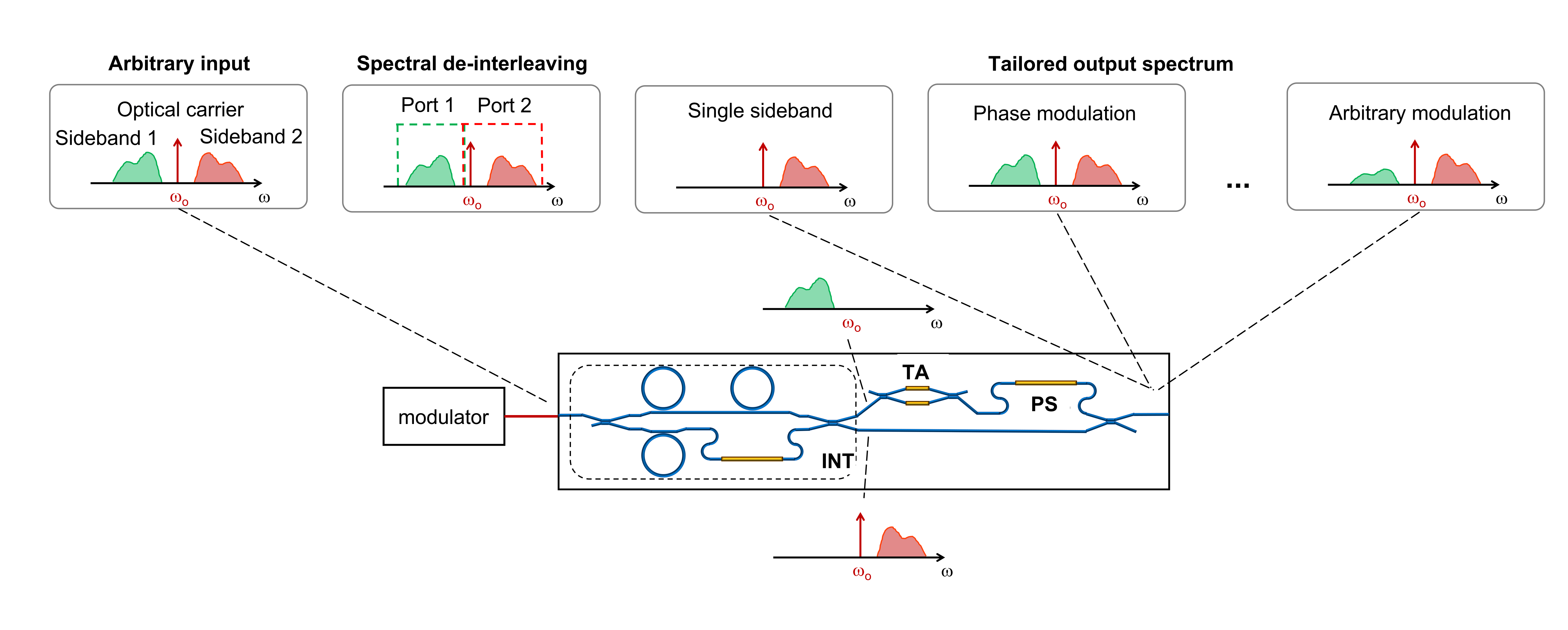}
\caption{\textbf{Operational principle of Modulation Transformer}. Schematic and operational principle of the modulation transformer (MT). The spectral de-interleaver is used to isolate one sideband from the optical carrier and the other sideband. A cascaded of tunable attenuator and a phase shifter is used to tailor the phase and amplitude of the isolated sideband. A combination with the optical carrier and unprocessed sideband leads to versatile spectral shaping, synthesizing variety of modulation format, such as single-sideband (SSB) modulation, phase modulation, and arbitrary modulation. INT: Spectral de-interleaver, TA: tunable attenuator, PS: phase shifter.}
\label{fig:figS1}
\end{figure}

\begin{figure}[b]
\centering
\includegraphics[width=\textwidth]{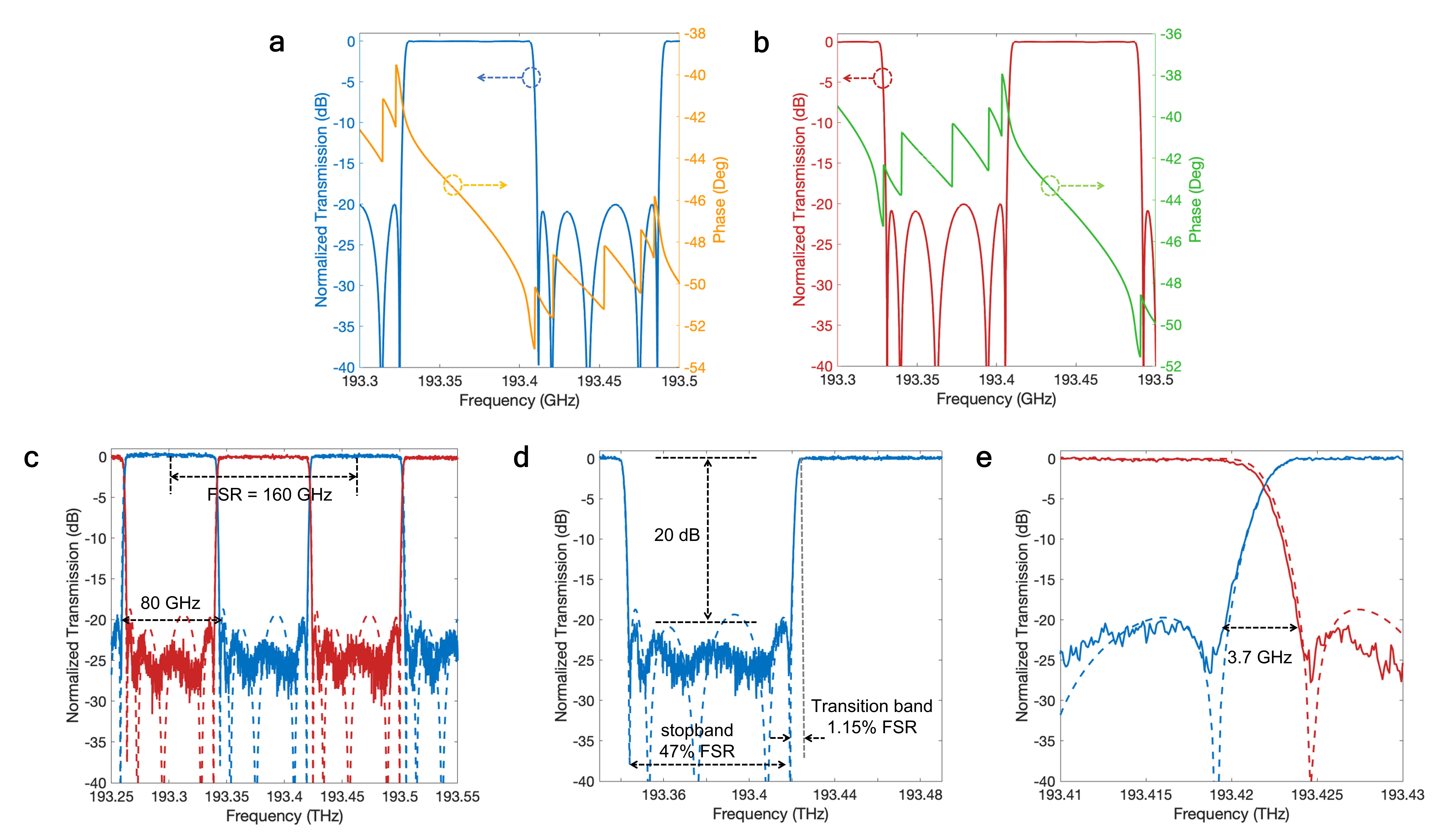}
\caption{\textbf{Spectral de-interleaver responses}. \textbf{a}, Simulation of amplitude and phase of the bar port from a spectral de-interleaver constructed using an asymmetric Mach-Zehnder Interferometer (aMZI) and three ring resonators. \textbf{b}, Simulation of amplitude and phase from the cross port of spectral de-interleaver. \textbf{c}, Fitting of simulation (dashed lines) and experimental results  of spectral de-interleaver.\textbf{ (d)} The rejection of spectral de-interleaver is 20~dB with stopband 47\% and transition band 1.15\% of spectral de-interleaver's FSR. \textbf{e}, The transition between bar and cross port of spectral de-interleaver is 3.7~GHz.}
\label{fig:figS2}
\end{figure}

A modulation transformer (MT) is an optical device designed to independently shape and synthesize arbitrary optical modulation spectrum. Such device constructed from multiple optical elements, namely spectral de-interleaver built from an asymmetric Mach-Zehnder Interferometer (aMZI) with three ring resonators topology \cite{luo2010high}, a tunable attenuator, a phase shifter and a combiner (Fig.~\ref{fig:figS1}). This device takes any conventional phase or intensity modulation spectrum as an input and synthesize an output optical modulation spectrum with components (optical carrier and sidebands) entirely independent in phases and amplitudes \cite{guo2021versatile}. This technique is different to prior techniques which mainly focus on PM-IM or IM-PM transformation by tailoring the phase of optical carrier.

The two output ports of spectral de-interleaver consist of one isolated sideband in one channel and an optical carrier with the remaining sideband in the other channel. Then, the isolated sideband is routed to cascaded phase shifter and tunable attenuator to independently tailor its phase and amplitude. Last, we re-combine this tailored isolated sideband with the rest of the optical spectrum component (optical carrier and unprocessed sideband) to synthesize different modulation spectrum with designer phase and amplitude relations between its spectrum elements. 

The mathematical description of each optical building block in the MT can be described as follows, 

First, a phase shifter is written as

\begin{equation}
\label{eq:eq1}
H_{ps}\left(z\right) = e^{-j \phi}
\vspace{-0.0cm}
\end{equation}

\noindent
then, a tunable coupler used for a tunable attenuator can be described as

\begin{equation}
\label{eq:eq2}
H_{tc}\left(z\right)=
\begin{bmatrix}
\sqrt{0.5} & -j \sqrt{0.5}\\
-j \sqrt{0.5} & \sqrt{0.5}
\end{bmatrix}
\begin{bmatrix}
1\\
e^{-j \phi}
\end{bmatrix}
\begin{bmatrix}
\sqrt{0.5} & -j \sqrt{0.5}\\
-j \sqrt{0.5} & \sqrt{0.5}
\end{bmatrix}
\vspace{-0.0cm}
\end{equation}

last, a spectral de-interleaver built from three ring resonators assisted aMZI structure can be expressed as

\begin{equation}
\label{eq:eq3}
H_{bar} = A_{11} A_{12} H_{11}\left(z\right) - A_{13} A_{14} H_{22} \left(z\right)
\end{equation}

\begin{equation}
\label{eq:eq4}
H_{cross} = A_{21} A_{22} H_{11}\left(z\right) - A_{23} A_{24} H_{22} \left(z\right)
\end{equation}

\noindent
where,

\begin{align*}
H_{11}\left(z\right) &= T_1\left(\omega\right) T_3\left(\omega\right)\\
H_{22}\left(z\right) &= T_2\left(\omega\right) H_{\Delta L} \left(z\right) H_{ps} \left(z\right)\\
H_{\Delta L} \left(z\right) &= \gamma z^{-1}\\
A_{11} &= c_3 c_4 e^{-j \phi_2} - s_3 s_4\\
A_{12} &= c_1 c_2 e^{-j \phi_1} - s_1 s_2\\
A_{13} &= c_4 s_3 e^{-j \phi_2} - c_3 s_4\\
A_{14} &= c_1 s_2 e^{-j \phi_1} - c_2 s_1\\
A_{21} &= c_3 c_4 e^{-j \phi_2} - s_3 s_4\\
A_{22} &= c_2 s_1 e^{-j \phi_1} - s_1 s_2\\
A_{23} &= c_3 c_4 - s_3 s_4 e^{-j \phi_2}\\
A_{24} &= c_1 s_2 e^{-j \phi_1} - c_2 s_1
\end{align*}

Here, c and s are the cross-coupling coefficient and the self-coupling coefficient of tunable element in the spectral de-interleaver respectively. Then, the amplitude response ($T\left(\omega\right)$) of the ring resonator used in the spectral de-interleaver can be expressed as \cite{bogaerts2012silicon}

\begin{equation}
\label{eq:eq5}
T\left(\omega\right)e^{-j \theta} = {\frac{a-c_r e^{-j \psi(\omega)}}{1-ac_r e^{-j \psi(\omega)}}}e^{-j (\pi + \psi(\omega))}
\end{equation}

\begin{figure}[ht]
\centering
\includegraphics[width=\textwidth]{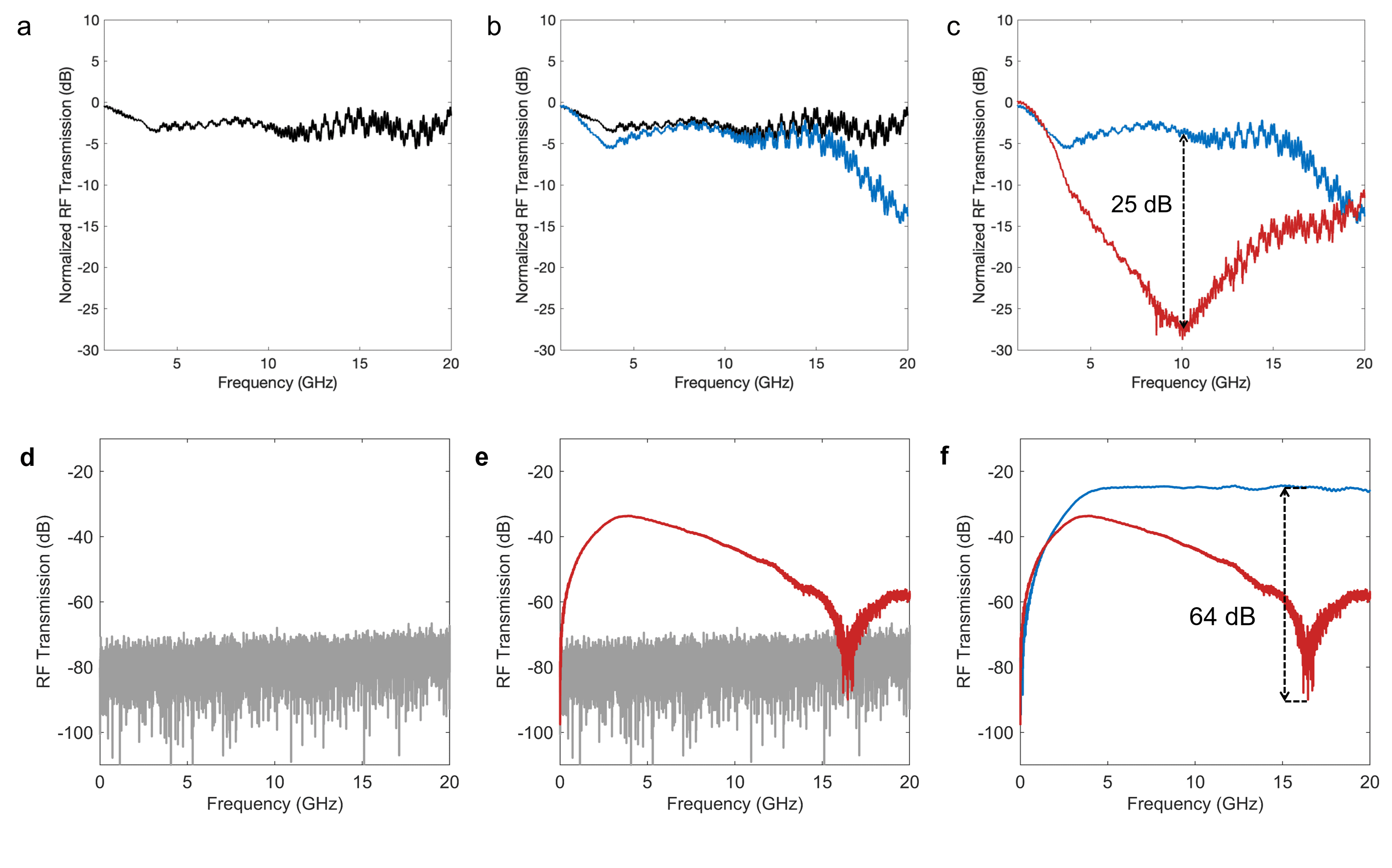}
\caption{\textbf{Experimental results of intensity-to-phase modulation (IM-PM) and phase-to-intensity modulation (PM-IM) conversions}. \textbf{a}, RF magnitude response of the intensity modulation (IM) link without the photonic integrated circuit (PIC) as a benchmark. \textbf{b},  RF magnitude response with the PIC containing MT and DI-RR (solid blue line). Here, the PIC response is set to be similar with the benchmark. \textbf{c}, Conversion from IM (blue) to phase modulation (PM) (red) with optimized extinction of 25~dB. \textbf{d}, RF magnitude response of the PM-based photonic link without the PIC as benchmark. \textbf{ (e)} RF magnitude response of the PM link with the PIC (solid red line). \textbf{f}, Conversion from PM (red) to IM (blue) with optimized extinction of 64~dB.}.
\label{fig:figS3}
\vspace{-0.1cm}
\end{figure}

\noindent
where $c_r=\sqrt{1-k}$, $a=10^{{-} \alpha L/20}$, k, $\alpha$, L, and $\psi$ are the self-coupling coefficient, the single-pass amplitude of the optical ring resonator, the coupling coefficient, the propagation loss of optical waveguide (dB/cm), the round trip length of ring, and the round-trip phase respectively. Similarly, we can define $s_r=\sqrt{k}$ as the cross-coupling coefficient.

Figure~\ref{fig:figS2}(a) and (b) show the simulated phase and amplitude responses from bar and cross port of spectral de-interleaver respectively. It is critical to synthesize a "box-shaped" response, flat-top to precisely isolate optical modulation spectrum elements (optical carrier and sidebands). The spectral de-interleaver has a linear phase response in the passband, while in the stopband the phase shows a nonlinear response, which may cause signal distortion in the MT. Figure~\ref{fig:figS2}(c) shows 160 GHz of spectral de-interleaver's free spectral range (FSR) with 80 GHz of bandwidth and rejection of up to 20 dB. The transition band in this response is $1.15\%$ of spectral de-interleaver's FSR or 1.85 GHz as shown in Fig.~\ref{fig:figS2}(d) with 3.7 GHz of transition bandwidth between bar and cross port as shown in Fig.~\ref{fig:figS2}(e). These results show the narrowest transition band reported in a spectral de-interlever device to date \cite{zhuang2011low}. The bar port then connected to a cascaded tunable attenuator and phase shifter. Last, a combiner at the output of the MT is used to re-combine the output of cascaded tunable attenuator and phase shifter with the output of cross port of spectral de-interleaver to synthesize different modulation spectrum.

\newpage
In this work, two examples are given to show modulation transformation process, such as intensity-to-phase modulation (IM-PM) transformation and phase-to-intensity modulation (PM-IM) transformation. Such modulation transformation can be mathematically described as

\begin{align*}
E_{up}\left(t\right) &= H_{tc}
\begin{bmatrix}
0\\
H_{ps} H_{11} E_{in}\left(t\right)
\end{bmatrix}\\
&= 0.5
\begin{bmatrix}
e^{-j\pi} - 1 & -j\left(e^{-j\pi} + 1\right)\\
-j\left(e^{-j\pi} + 1\right) & 1 - e^{-j\pi}
\end{bmatrix}
\begin{bmatrix}
0\\
H_{ps} H_{11} E_{in}\left(t\right)
\end{bmatrix}\\
&=
\begin{bmatrix}
-1 & 0\\
0 & 1
\end{bmatrix}
\begin{bmatrix}
0\\
H_{ps} H_{11} E_{in}\left(t\right)
\end{bmatrix}\\
&=
\begin{bmatrix}
0\\
H_{ps} H_{11} E_{in}\left(t\right)
\end{bmatrix}\\
&= H_{ps} H_{11} E_{in}\left(t\right)\\
&= e^{-j\pi} H_{11} E_{in}\left(t\right)\\
&= -H_{11} E_{in}\left(t\right)\\
E_{low}\left(t\right) &= H_{12} H_{\Delta L1} E_{in}\left(t\right)\\
E_{comb}\left(t\right) &= E_{up}\left(t\right) + e^{-j \pi /2} E_{low}\left(t\right)\\
&= -\left(H_{11} + j H_{\Delta L1} H_{12}\right) E_{in}\left(t\right)
\end{align*}

\noindent
where $E_{up}\left(t\right)$ and $E_{low}\left(t\right)$ are the output field of upper and lower sidebands respectively. 

We controlled the MT in our circuit to synthesis two different modulation schemes from a conventional PM and IM input. In the first experiment, we sent an IM spectrum directly to a photodetector in a simple photonic link setup and measured the RF transmission for a reference. Then, we process the IM spectrum using the MT in our circuit and convert into PM spectrum. Figure~\ref{fig:figS3}(a) shows the IM spectrum in the RF domain as the reference signal in a simple photonic link setup. Then, the IM spectrum is sent to our circuit containing of a MT as shown in Fig.~\ref{fig:figS3}(b). Here, we try to match the RF spectrum (with photonic chip) with the reference spectrum (without photonic chip). While both of the sidebands in IM spectrum are in-phase, the measured RF transmission is high (solid blue line in Fig.~\ref{fig:figS3}(b) and (c)). Later, we tailored the phase of isolated sideband by tuning the phase shifter in the MT. Last, the processed isolated sideband is re-combined with the optical carrier and remaining sideband to synthesized the PM spectrum (solid red line in Fig.~\ref{fig:figS3}(c)) with 25 dB extinction.

Next, in second experiment, we used the MT for PM-IM transformation with the same principle as prior experiment, but with the PM spectrum as an input. Figure~\ref{fig:figS3}(d) shows the transmission of PM spectrum as a benchmark in a simple photonic link setup with direct detection. Then, we sent the PM spectra into the un-activated MT in our circuit as shown in Fig.~\ref{fig:figS3}(e). When both sidebands are out of phase, the measured RF transmission is low (solid red line in Fig.~\ref{fig:figS3}(e) and (f)). Then we tuned the phase shifter in the MT to rotate the phase of isolated sideband. Eventually, an IM spectrum is synthesized after the signal recombination, and it is corresponding to a maximum RF transmission (solid blue line in Fig.~\ref{fig:figS3}(f)) with 64 dB extinction. In these two experiments, the operational bandwidth and extinction are mainly limited by the roll-off and the dispersion of the spectral de-interleaver, notably at the transition band (operational frequency of $<$ 5 GHz).

\newpage
\section*{Supplementary Information B: Double Injection Ring Resonator}

\begin{figure}[ht]
\centering
\includegraphics[width=\textwidth]{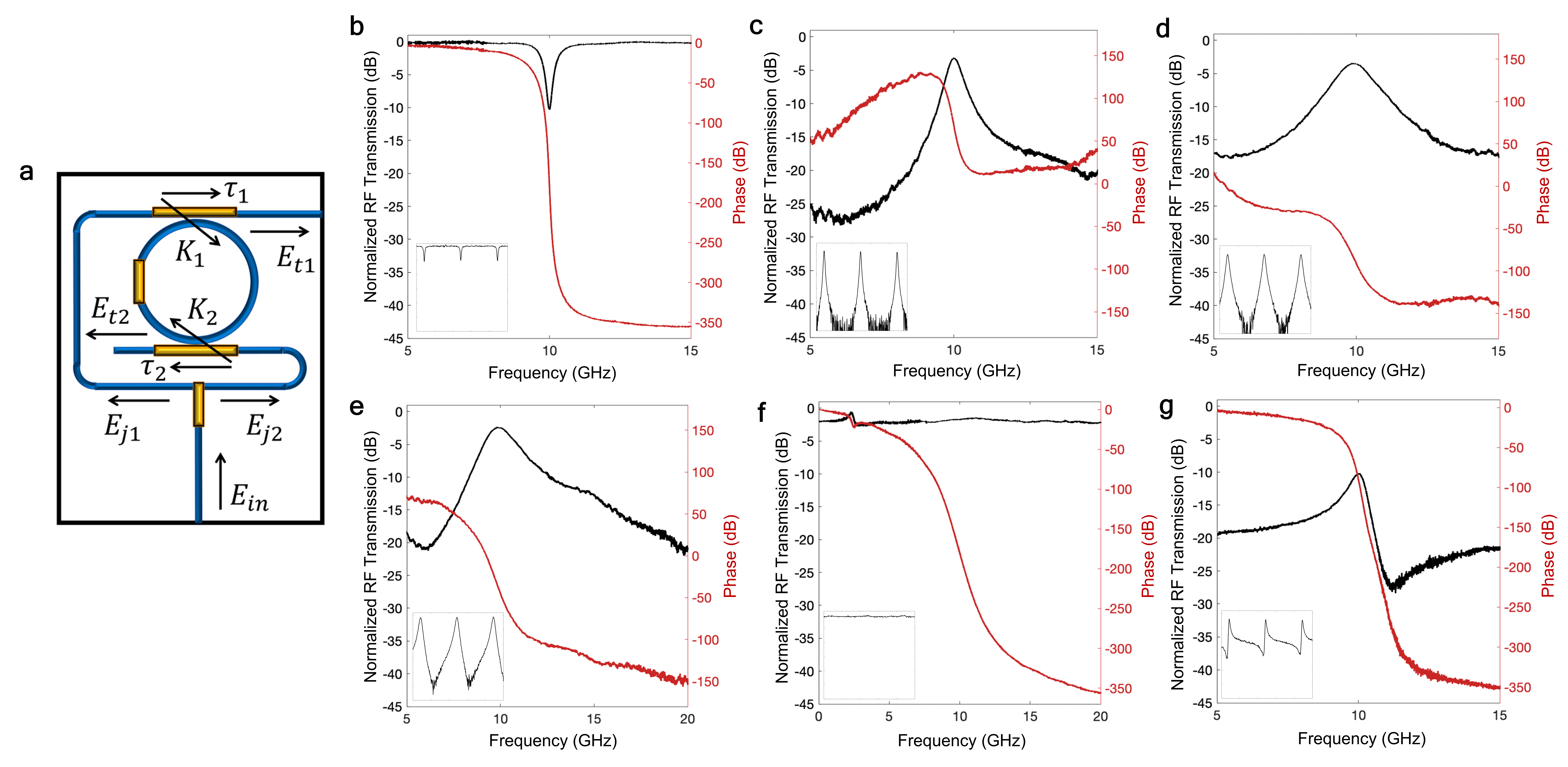}
\caption{\textbf{RF phase and magnitude responses of the DI-RR}. \textbf{a}, Schematic of double injection ring resonator. \textbf{b}, Notch filter. \textbf{(c}) Bandpass filter. \textbf{d}, Triangular response \textbf{e}, Sawtooth response. \textbf{f}, All-pass response. \textbf{g}, Fano-like response. The inset of each figure shows the corresponding optical response.}
\label{fig:figS4}
\vspace{-0.1cm}
\end{figure}

The use of an optical circuit named double-injection ring resonator (DI-RR) is aimed to provide various unique responses in a ring resonator-based optical circuit from single output. The underlying idea of this circuit is to inject two, mutually coherent, optical signals of the same wavelength into a single add-drop ring resonator from single input \cite{cohen2018response} as illustrated in Fig.~\ref{fig:figS4}(a). To optimize such circuit for our device, we further investigate the variety of unique phase and amplitude response synthesized by the DI-RR, which is a key parameter for programmable integrated MWP circuit. 

Mathematically, the model describing the transmitted electric field dependence on the wavelength of the DI-RR is given by

\begin{equation}
\label{eq:eq6}
E_{t1}\left(\lambda\right)=\frac{\left(\tau_1 + \tau_2^{*} \alpha e^{-j \theta}\right)}{1 - \tau_1 \tau_2^{*} \alpha e^{-j \theta}} \big|E_{j1}\left(\lambda\right)\big|e^{-j \phi_{j1}} - \frac{K_1 K_2^{*} \sqrt{\alpha} e^{-j \theta}}{1 - \tau_1 \tau_2^{*} \alpha e^{-j \theta}} \big|E_{j2}\left(\lambda\right) \big|e^{-j \phi_{j2}}
\end{equation}

\noindent
where $\tau = \big|\tau\big|e^{-j \psi_\tau}$ is the transmission of the directional coupler, $K = \big| K \big| e^{-j \psi_K}$ is the coupling coefficients of the directional couplers, $\alpha$ is the loss coefficient of the ring, $E_j$ is the injected fields, and $\phi_j$ is the injected fields phases. $\theta$ is the phase accumulated by the light traversing the ring at steady state that described as

\begin{equation}
\label{eq:eq7}
\theta\left(\lambda\right)=\frac{2 \pi}{\lambda} n_{eff} \left(\lambda\right) L_{ring}
\end{equation}

\noindent
with $\lambda$ being the wavelength, $L_{ring}$ the length perimeter of the ring, and $n_{eff}$ the effective index of the propagating mode. Fig.~\ref{fig:figS4}(b--g) depict the measured RF phase and amplitude responses of the six different functions synthesized from the DI-RR, such as notch filter (Fig.~\ref{fig:figS4}(b)), bandpass filter (Fig.~\ref{fig:figS4}(c)), triangular response (Fig.~\ref{fig:figS4}(d)), sawtooth response (Fig.~\ref{fig:figS4}(e)), all-pass response (Fig.~\ref{fig:figS4}(f)) and Fano-like response (Fig.~\ref{fig:figS4}(g)). The inset of each figure depict the response obtained from high resolution optical spectrum analysis. The characterization is important to have a full understanding of the DI-RR's potential in the system.

\newpage
\section*{Supplementary Information C: Extended Experiments}

\begin{figure*}[ht]
\centering
\includegraphics[width=0.8\textwidth]{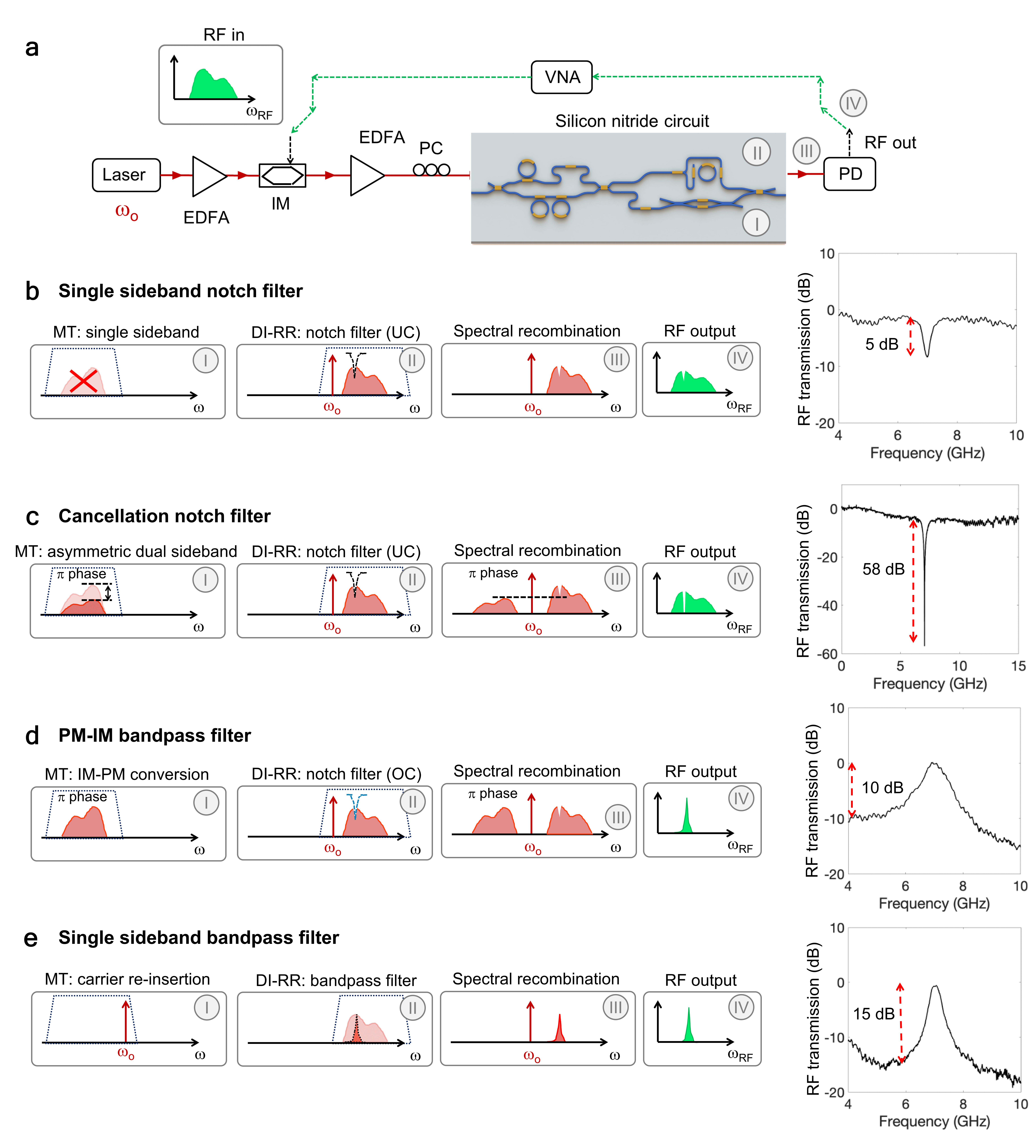}
\caption{\small \textbf{Experiment scenarios of programmable integrated MWP circuit.} \textbf{(a)} Experiment setup used for demonstrating the four different filtering scenarios. The modulation transformer (MT) and the double-injection ring resonator (DI-RR) are programmed to exhibit different functions using a common intensity modulation (IM) input.  \textbf{(b)} Scenario 1: Single sideband notch filter. The MT is used to synthesize single sideband modulation (SSB) and the DI-RR to synthesize 5~dB-deep notch response. The measured filter response is show on the right panel. \textbf{(c)} Scenario 2: Cancellation notch filter. MT is used to create asymmetric dual-sideband modulation conversion while DI-RR shows 5 dB-deep notch response. Destructive interference at the otch frequency amplifies the RF notch filter response to 58 dB. \textbf{(d)} Scenario 3: PM-IM bandpass filter. The MT is used to create phase modulation.  The over-coupled (OC) notch response from DI-RR breaks the PM condition and creates a bandpass filter response.\textbf{ (e)} Scenario 4: Single-sideband bandpass filter. The MT is used to re-insert an optical carrier to mix with a single sideband response that has been filtered using the DI-RR. The DI-RR was tuned to exhibit a sharp bandpass filter response.}
\label{fig:figS5}
\end{figure*}

\newpage
To demonstrate the feasibility of the MT together with the versatility of DI-RR in programmable integrated MWP circuit, we conducted experiments with an intensity modulation (IM) spectrum as the input optical spectrum. Four different scenarios were applied to demonstrated the versatility of the circuit. Figure~\ref{fig:figS5}(a) shows the experiment setup and Fig.~\ref{fig:figS5}(b - e) show the operation principle and application of four different scenarios of the proposed device. In the system, the input RF signal is upconverted to optical domain with intensity modulator generating single frequency optical carrier and two identical amplitude sidebands that are in-phase.

In the first scenario, we convert an intensity-to-single sideband (SSB) modulation to, later on, create a SSB RF notch filter as shown in Fig.~\ref{fig:figS5}(b). An IM signal is sent to pass through a spectral de-interleaver in the MT, which spatially isolated one sideband in the spectrum from another sideband and optical carrier. Then, the isolated sideband was sent to tunable attenuator to be fully attenuated, where the other sideband and optical carrier were sent to a DI-RR which synthesized a notch response. Last, the output of DI-RR was combined with the fully attenuated sideband and exhibited a SSB RF notch filter. 

\begin{figure*}[ht]
\centering
\includegraphics[width=0.85\textwidth]{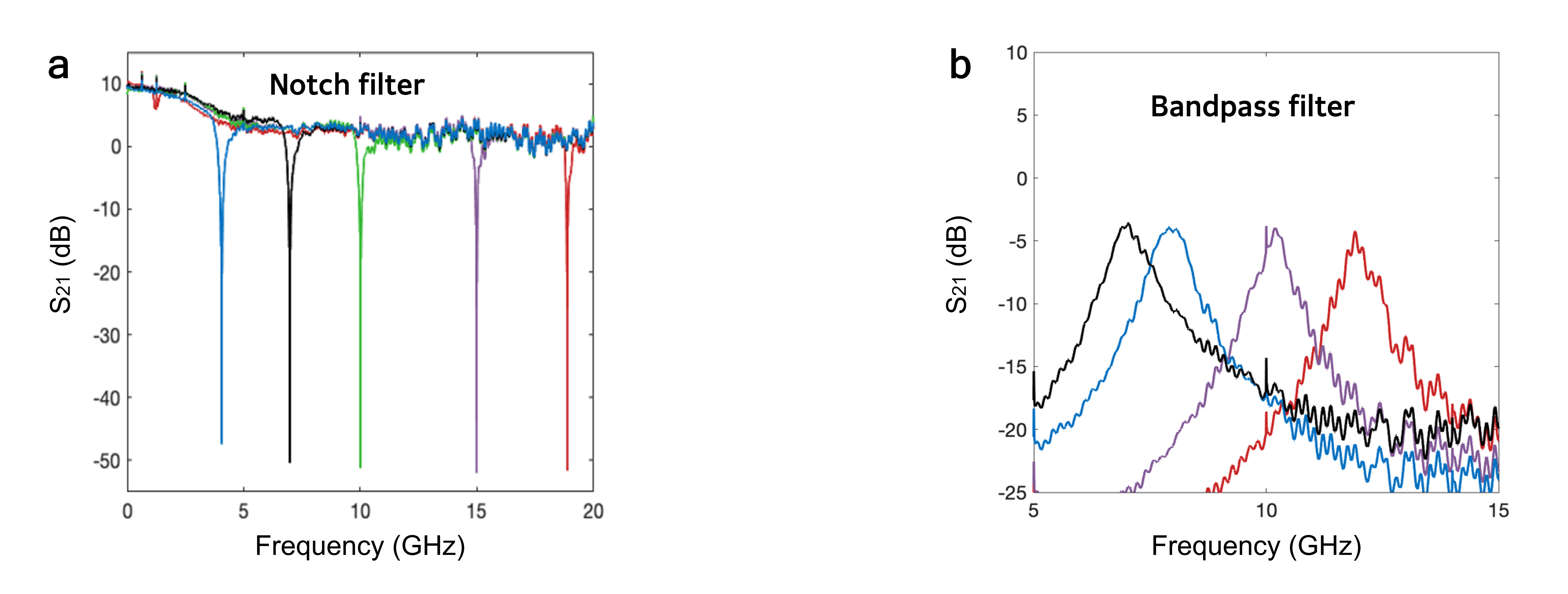}
\caption{\small \textbf{RF Filters Tunability.} \textbf{(a)} Tuning range of RF notch filter with low-biasing MZM. \textbf{(b)} Tuning range of RF bandpass filter with low-biasing MZM.}
\label{fig:figS6}
\vspace{-0.1cm}
\end{figure*}

Then, in the second scenario, an IM signal is sent to the MT, perform intensity-to-asymmetric dual sideband (aDSB) modulation conversion and synthesize high rejection RF notch filter. Here, the attenuation of isolated sideband is similar with the rejection of notch response synthesized by the DI-RR. Then, the phase shifter was used to create phase difference (0 - $\pi$) between two sidebands relative to optical carrier. In parallel, the other sideband and optical carrier were processed by the same notch response as in the first scenario. It is important that the rejection of notch response from the ring to be equal with the amplitude’s difference between two sidebands. The idea is to have a desctructive interference between aDSB together with notch response from DI-RR to amplify the rejection of synthesized RF notch filter as shown in Fig.~\ref{fig:figS5}(c). By precisely tailor the phase of the DI-RR, we can tune the central frequency of the RF notch filter as shown in Fig.~\ref{fig:figS6}(a).

Next, in the third scenario, we conducted phase-to-intensity modulation (PM-IM) conversion technique to create RF bandpass filter. Here, we converted the modulation spectrum from intensity-to-phase modulation (IM-PM). During the process, we only changed the phase of the isolated sideband by controlling the interconnected phase shifter from 0 - $\pi$. While, the other sideband and optical carrier were processed using notch response of DI-RR set at over-coupling (OC) regime. After IM-PM conversion using MT, due to OC state in the DI-RR, a $\pi$-phase shift is introduced at the desired notch frequency, creating constructive interference for phase-to-intensity (PM-IM) modulation conversion and synthesized RF bandpass filter as shown in Fig.~\ref{fig:figS5}(d). Previously, this approach can only be done in a phase modulator-based MWP system.

Last, another RF bandpass filter is exhibited using optical carrier re-insertion technique with the MT. In this approach, a SSB modulation is sent to the spectral de-interleaver in the MT and split into two outputs containing only optical carrier in one output and one sideband in the other. Then, the sideband was sent to a DI-RR which tuned and synthesized a bandpass response. The RF bandpass filter is constructed while the optical carrier is re-injected to the processed sideband using a combiner at the output of MT as shown in Fig.~\ref{fig:figS5}(e). Figure~\ref{fig:figS6}(b) shows the tuning range of the RF bandpass filter achieved by tailoring the phase of DI-RR.

\newpage
\section*{Supplementary Information D: Theoretical Analysis of Linearization Method}

A theoretical analysis is performed to find the requirements for linearity improvement in the proposed RF notch filter. We first analyze the relation between optical spectrum and the beating products after photodetection to find the main contributors of the IMD3. When the optical carrier is modulated by a two-tone RF signal with angular frequency of $\omega_1$, $\omega_2$, and voltage of $V_{RF}$ via a phase modulator (PM), the optical spectrum at the output of PM can be expressed as

\begin{equation}
\label{eq:eq8}
E_{out}\left(t\right)\hspace{-0.06cm}=\hspace{-0.1cm}\sqrt{P_i}e^{j\omega_{c}t}\hspace{-0.1cm}\sum_{\scriptscriptstyle n=\hspace{-0.02cm}-\hspace{-0.02cm}\infty}^{\scriptscriptstyle +\infty}\hspace{-0.05cm}\sum_{\scriptscriptstyle k=\hspace{-0.02cm}-\hspace{-0.02cm}\infty}^{\scriptscriptstyle+\infty} \hspace{-0.15cm}J_{n}\left(m\right)\hspace{-0.05cm}J_{k}\left(m\right)\hspace{-0.05cm}e^{j\left(n\omega_{1}+k\omega_{2}\right)t} 
\vspace{-0.0cm}
\end{equation}

\noindent
where $\omega_{c}$, $P_{i}$, $J_{n}$, $m=\pi \cdot V_{RF}/V_{\pi,RF}$, and $V_{\pi,RF}$ is the angular frequency of the optical carrier, input optical power, the n-th order Bessel function of the first kind, the modulation index of PM, and the RF half-wave voltage of the PM respectively. When the two-tone signal can be regarded as small signal, we can only take zero to second order sidebands into account to simplify the analysis.

The optical spectrum after spectral shaping can be written as
\begin{equation}
\label{eq:eq9}
\begin{aligned}
E_{p}\left(t\right)\hspace{-0.1cm}&=\hspace{-0.1cm}\sqrt{P_i}e^{j\omega_{c}t}\\
&\cdot\left\{\hspace{-0.2cm}
\begin{array}{ll}
     \sqrt{A}\cdot J_{0}J_{0}\vspace{1ex}\\
     +\sqrt{A}\cdot J_{-1}J_{1}\left[e^{-j\left(\omega_{1}+\omega_{2}\right)t}+e^{j\left(\omega_{1}+\omega_{2}\right)t}\right]\vspace{1ex} \\ 
     +J_{0}J_{1}\left(e^{j\omega_{1}t}+e^{j\omega_{2}t}\right)\vspace{1ex}\\
     +J_{-1}J_{2}\left[e^{j\left(2\omega_{1}-\omega_{2}\right)t}+e^{j\left(2\omega_{2}-\omega_{1}\right)t}\right]\vspace{1ex} \\
     +\sqrt{A}\cdot J_{-1}J_{0}\left(e^{-j\omega_{1}t}+e^{-j\omega_{2}t}\right)\vspace{1ex}\\
     +\sqrt{A}\hspace{-0.08cm}\cdot\hspace{-0.08cm} J_{-2}J_{1}\left[e^{-j\left(2\omega_{1}-\omega_{2}\right)t}\hspace{-0.1cm}+\hspace{-0.1cm}e^{-j\left(2\omega_{2}-\omega_{1}\right)t}\right] \vspace{1ex}\\ 
     +J_{0}J_{2}\left(e^{j2\omega_{1}t}+e^{j2\omega_{2}t}\right)\vspace{1ex}\\
     +J_{1}J_{1}e^{j\left(\omega_{1}+\omega_{2}\right)t}\vspace{1ex}\\
     +\sqrt{A}\cdot J_{-2}J_{0}\left(e^{-j2\omega_{1}t}+e^{-j2\omega_{2}t}\right)\vspace{1ex}\\
     +\sqrt{A}\cdot J_{-1}J_{-1}e^{-j\left(\omega_{1}+\omega_{2}\right)t}\\
\end{array}\hspace{-0.2cm}\right\}
\end{aligned}
\end{equation}

\noindent
where $J_{n}=J_{n}(m) (n=0, \pm1, \pm2)$, $J_{-n}=(-1)^{n}J_{n}$. A is the power suppression to the optical carrier and the lower sideband. 

The photocurrent of the RF signal detected from the processed optical spectrum can be expressed as

\begin{equation}
\label{eq:eq10}
\begin{aligned}
I_{PD}(t)&=R_{PD}|E_{p}(t)|^{2}\\
&=I_{1}\cos{\omega_{1,2}t}+I_{3}\cos{\left(2\omega_{1,2}-\omega_{2,1}\right)t}
\end{aligned}
\end{equation}

\noindent
where $R_{PD}$ is responsivity of photodetector, $I_{1}$ and $I_{3}$ are the amplitude coefficients for fundamental signal and IMD3 components, which can be written as

\begin{equation}
\label{eq:eq11}
\begin{aligned}
I_{1} \propto(\sqrt{A}\hspace{-0.07cm}-\hspace{-0.07cm}A) J_{0}^{3}J_{1}+(1\hspace{-0.07cm}-\hspace{-0.07cm}\sqrt{A})J_{0}J_{1}^{3}+(1\hspace{-0.07cm}-\hspace{-0.07cm}A)J_{0}^{2}J_{1}J_{2}
\end{aligned}
\end{equation}

\begin{equation}
\label{eq:eq12}
\begin{aligned}
I_{3} \propto(A-\sqrt{A}) J_{0} J_{1}^{3}+(1-\sqrt{A}) J_{0}^{2} J_{1} J_{2} 
\end{aligned}
\vspace{-0.0cm}
\end{equation}

\noindent
when $m\ll1$, $J_{n}(m)\approx m^{n}/(2^{n}n!)$, $I_{1}$ and $I_{3}$ can be expressed as

\begin{equation}
\label{eq:eq13}
\begin{aligned}
I_{1} \propto(\sqrt{A}-A) m 
\end{aligned}
\end{equation}

\begin{equation}
\label{eq:eq14}
\begin{aligned}
I_{3} \propto(2 A-3 \sqrt{A}+1) m^{3}
\end{aligned}
\end{equation}

\newpage
\section*{Supplementary Information E: Operational Principle of Simultaneous Filter and Linearization}

\begin{figure*}[ht]
\vspace{-0.5cm}
\centerline{\includegraphics[width=0.85\textwidth]{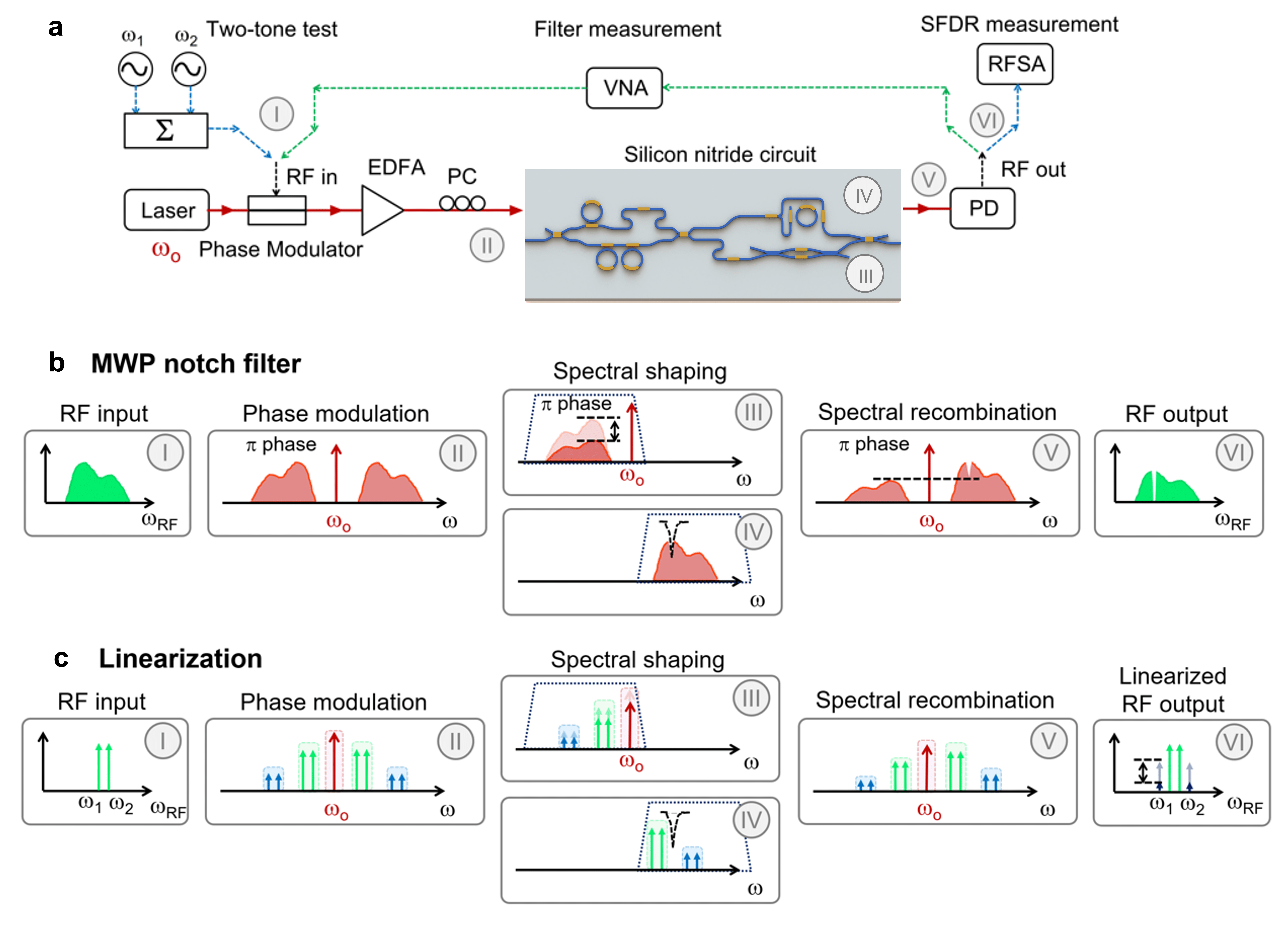}}
\caption{\small \textbf{Linearization of RF notch filter.}\textbf{(a)} Experiment setup used for demonstrating simultaneous RF notch filter with linearization. \textbf{(b)} Signal flow for RF notch filter formation. The filter is formed through RF cancellation at the notch frequency. \textbf{(c)} Signal flow for the simultaneous RF notch filter and linearization. The amplitude and phase of the optical carrier and first and second order sidebands are adjusted to achieve cancellation of IMD3 products at the output.}
\label{fig:figS7}
\end{figure*}

In this experiment, we experimentally demonstrated, for the first time, simultaneous RF notch filter and linearization using a combination of the MT and the DI-RR in a single photonic chip. The DI-RR is set to create a notch response, while The MT works for two purposes, synthesize desired modulations spectrum needed for RF notch filter and linearization. We conducted the experiment with setup as depicted in Fig.~\ref{fig:figS7}(a) with the operating principle of simultaneous RF notch filter and linearization as illustrated in Fig.~\ref{fig:figS7}(b) and (c).

The RF notch filter here is synthesized with the basis of phase cancellation \cite{marpaung2015low}. First, the RF signal (\RomanNumeralCaps{1}) is phase modulated into the optical carrier generating single frequency optical carrier with multiple sidebands that are out-of-phase (\RomanNumeralCaps{2}). The phase modulation spectrum then coupled into the programmable integrated MWP circuit containing of the MT and DI-RR. The spectral de-interleaver in the MT spatially separates the phase modulation spectrum into two parts, where the optical carrier and lower sideband are attenuated by the tunable attenuator (\RomanNumeralCaps{3}). The upper sideband is processed by notch response from the DI-RR set at under-coupling (UC) regime to equalize the amplitudes of two sidebands at designate frequency (\RomanNumeralCaps{4}). Next, the optical spectrum are recombined at the output of the chip, creating an asymmetric dual sideband (aDSB) modulation with anti-phase relation (\RomanNumeralCaps{5}). After photodetection, a high rejection RF notch filter is achieved, as the mixing products of the sidebands and the optical carrier are destructively interfered at the notch frequency (\RomanNumeralCaps{6}).

Next, the operating principle of simultaneous RF notch filter and linearization can be explained as, two-tone RF signal located at the passband of the RF notch filter (\RomanNumeralCaps{1}) is phase modulated into the optical carrier. Here, the multi-order sidebands in phase-modulated spectrum are taken into account for IMD3 reduction (\RomanNumeralCaps{2}). Then, the optical spectrum is coupled to the photonic chip, where the spectral de-interleaver then split the optical modulation spectrum into two channels. one channel contains the optical carrier, the -1 order and -2 order sidebands, which are manipulated by a cascaded tunable attenuator and phase shifter to create aDSB modulation for RF notch filter and meet the linearization condition (\RomanNumeralCaps{3}). The other channel consists of the +1 order and +2 order sidebands. The +1 order sideband is processed by a notch response from the DI-RR (\RomanNumeralCaps{4}). As the two-tone signal lies in the passband and the DI-RR is used to tailor the frequency components at stopband, the DI-RR's response will not influence much to the linearity. Then, the manipulated spectrum are recombined at the output of the chip (\RomanNumeralCaps{5}) and sent to the photodetector, resulting in an RF spectrum with suppressed IMD3 terms due to the destructive interference between the mixing products of the multi-order sidebands and the optical carrier (\RomanNumeralCaps{6}).

\end{document}